\input harvmac
\input epsf
\input amssym
%\draftmode
%
%
\noblackbox
%%%%%%%%%%%%%%%%%%%%%%%%%%%%%%%%%%%%%%%%%%%%%%
%%%%%%%%%%%%%%%%%%%%%%%%%%%
% some stuff needed for figures:
%%%%%%%%%%%%%%%%%%%%%%%%%%%%%%%%%%%%%%%%%%%%%%
%%%%%%%%%%%%%%%%%%%%%%%%%%%
\newcount\figno
\figno=0
\def\fig#1#2#3{
\par\begingroup\parindent=0pt\leftskip=1cm\rightskip=1cm\parindent=0pt
\baselineskip=11pt
\global\advance\figno by 1
\midinsert
\epsfxsize=#3
\centerline{\epsfbox{#2}}
\vskip -21pt
{\bf Fig.\ \the\figno: } #1\par
\endinsert\endgroup\par
}
\def\figlabel#1{\xdef#1{\the\figno}}
\def\encadremath#1{\vbox{\hrule\hbox{\vrule\kern8pt\vbox{\kern8pt
\hbox{$\displaystyle #1$}\kern8pt}
\kern8pt\vrule}\hrule}}
%%%%%%%%%%%%%%%%%%%%%%%%%%%%%%%%%%%%%%%%%%%%%%
%%%%%%%%%%%%%%%%%%%%%%%%%%%
% definitions
%%%%%%%%%%%%%%%%%%%%%%%%%%%%%%%%%%%%%%%%%%%%%%
%%%%%%%%%%%%%%%%%%%%%%%%%%%

\def\frac#1#2{{#1 \over #2}}

\def\p{\partial}
\def\semi{\subset\kern-1em\times\;}
\def\bar#1{\overline{#1}}
\def\sqr#1#2{{\vcenter{\vbox{\hrule height.#2pt
\hbox{\vrule width.#2pt height#1pt \kern#1pt \vrule width.#2pt}
\hrule height.#2pt}}}}

\def\p{\partial}

\def\ad{\bar a}

\def\p{\partial}

\def\ct{\tilde{c}}

\def\Nc{{\cal N}}
\def\Dc{{\cal D}}

\def\ct{\tilde{c}}

\def\ih{\hat{i}}
\def\jh{\hat{j}}
\def\kh{\hat{k}}
\def\lh{\hat{l}}

\def\Mh{\hat{M}}
\def\Lc{{\cal L}}

%
%%%%%%%%%%%%%%%%%%%%%%%%%%%%%%%%%%%%%%%%%%%%%%
%%%%%%%%%%%%%%%%%%%%%%%%%%%
% more definitions
%%%%%%%%%%%%%%%%%%%%%%%%%%%%%%%%%%%%%%%%%%%%%%
%%%%%%%%%%%%%%%%%%%%%%%%%%%

%
%\def\oneone{\rlap 1\mkern4mu{\rm l}}
%\def\coeff#1#2{\relax{\textstyle {#1 \over #2}}\displaystyle}

\def\IR{\Bbb{R}}

\def\zb{\overline{z}}

\def\Nc{{\cal N}}

\def\Mh{\hat{M}}
\def\jh{\hat{j}}

%%%%%%%%%%%%%%%%%%%%%%%%%%%%%%%%%%%%%%%%%%%%%%
%%%%%%%%%%%%%%%%%%%%%%%%%%%
% References
%%%%%%%%%%%%%%%%%%%%%%%%%%%%%%%%%%%%%%%%%%%%%%
%%%%%%%%%%%%%%%%%%%%%%%%%%%

\lref\masakiref{
%\IizukaUV
  N.~Iizuka and M.~Shigemori,
  ``A Note on D1-D5-J System and 5D Small Black Ring,''
  [arXiv:hep-th/0506215].
  %%CITATION = HEP-TH 0506215;%%
}

  %\KrausGH
\lref\KLatt{
  P.~Kraus and F.~Larsen,
  ``Attractors and black rings,''
  Phys.\ Rev.\ D {\bf 72}, 024010 (2005)
  [arXiv:hep-th/0503219].
  %%CITATION = HEP-TH 0503219;%%
}

%\ChamseddinePI
\lref\ChamseddinePI{
  A.~H.~Chamseddine, S.~Ferrara, G.~W.~Gibbons and R.~Kallosh,
  ``Enhancement of supersymmetry near 5d black hole horizon,''
  Phys.\ Rev.\ D {\bf 55}, 3647 (1997)
  [arXiv:hep-th/9610155].
  %%CITATION = HEP-TH 9610155;%%
}

\lref\StromBTZ{ A.~Strominger,
 ``Black hole entropy from near-horizon microstates'',
JHEP {\bf 9802}, 009 (1998); [arXiv:hep-th/9712251];
V.~Balasubramanian and F.~Larsen, ``Near horizon geometry and black
holes in four dimensions'', Nucl.\ Phys.\ B {\bf 528}, 229 (1998);
[arXiv:hep-th/9802198].
%%CITATION = HEP-TH 9802198;%%
}

%\BalasubramanianEE
\lref\BalasubramanianEE{
  V.~Balasubramanian and F.~Larsen,
  %``Near horizon geometry and black holes in four dimensions,''
  Nucl.\ Phys.\  B {\bf 528}, 229 (1998)
  [arXiv:hep-th/9802198].
  %%CITATION = NUPHA,B528,229;%%
}

\lref\MSW{ J.~M.~Maldacena, A.~Strominger and E.~Witten, ``Black
hole entropy in M-theory'', JHEP {\bf 9712}, 002 (1997);
[arXiv:hep-th/9711053]. }
  %%CITATION = HEP-TH 9711053;%%

\lref\HMM{ J.~A.~Harvey, R.~Minasian and G.~W.~Moore, ``Non-abelian
tensor-multiplet anomalies,''
 JHEP {\bf 9809}, 004 (1998)
  [arXiv:hep-th/9808060].
 %%CITATION = HEP-TH 9808060;%%
}

\lref\antRRRR{  I.~Antoniadis, S.~Ferrara, R.~Minasian and
K.~S.~Narain,
  ``R**4 couplings in M- and type II theories on Calabi-Yau spaces,''
  Nucl.\ Phys.\ B {\bf 507}, 571 (1997)
  [arXiv:hep-th/9707013].
  %%CITATION = HEP-TH 9707013;%%
 }

\lref\WittenMfive{ E.~Witten,
  ``Five-brane effective action in M-theory,''
  J.\ Geom.\ Phys.\  {\bf 22}, 103 (1997)
  [arXiv:hep-th/9610234].
  %%CITATION = HEP-TH 9610234;%%
}

\lref\wittenAdS{ E.~Witten,
  ``Anti-de Sitter space and holography,''
  Adv.\ Theor.\ Math.\ Phys.\  {\bf 2}, 253 (1998)
  [arXiv:hep-th/9802150].
  %%CITATION = HEP-TH 9802150;%%
  }

\lref\brownhen{  J.~D.~Brown and M.~Henneaux,
 ``Central Charges In The Canonical Realization Of Asymptotic Symmetries: An
  Example From Three-Dimensional Gravity,''
  Commun.\ Math.\ Phys.\  {\bf 104}, 207 (1986).
  %%CITATION = CMPHA,104,207;%%
  }

\lref\wald{
  R.~M.~Wald,
  ``Black hole entropy is the Noether charge,''
  Phys.\ Rev.\ D {\bf 48}, 3427 (1993)
  [arXiv:gr-qc/9307038].
  %%CITATION = GR-QC 9307038;%%
R.~Wald, Phys.\ Rev.\ D {\bf 48} R3427 (1993);
   V.~Iyer and R.~M.~Wald,
  ``Some properties of Noether charge and a proposal for dynamical black hole
  entropy,''
  Phys.\ Rev.\ D {\bf 50}, 846 (1994)
  [arXiv:gr-qc/9403028].
  %%CITATION = GR-QC 9403028;%%
 ``A Comparison of Noether charge and Euclidean methods for computing the
  entropy of stationary black holes,''
  Phys.\ Rev.\ D {\bf 52}, 4430 (1995)
  [arXiv:gr-qc/9503052].
  %%CITATION = GR-QC 9503052;%%
}

\lref\senrescaled{A.~Sen,
   ``Black holes, elementary strings and holomorphic anomaly,''
     JHEP {\bf 0507}, 063 (2005)
  [arXiv:hep-th/0502126];
  %%CITATION = HEP-TH 0502126;%%;
   ``Entropy function for heterotic black holes,''
      JHEP {\bf 0603}, 008 (2006)
  [arXiv:hep-th/0508042]
  %%CITATION = JHEPA,0603,008;%%
;
  %\SahooPM
  B.~Sahoo and A.~Sen,
  ``alpha' corrections to extremal dyonic black holes in heterotic string
  theory,''
  JHEP {\bf 0701}, 010 (2007)
  [arXiv:hep-th/0608182].
  %%CITATION = JHEPA,0701,010;%%
}

  %\SenWA
\lref\SenWA{
  A.~Sen,
  ``Black hole entropy function and the attractor mechanism in higher
  derivative gravity,''
  JHEP {\bf 0509}, 038 (2005)
  [arXiv:hep-th/0506177].
  %%CITATION = JHEPA,0509,038;%%
}

\lref\saidasoda{
  H.~Saida and J.~Soda,
  ``Statistical entropy of BTZ black hole in higher curvature gravity,''
  Phys.\ Lett.\ B {\bf 471}, 358 (2000)
  [arXiv:gr-qc/9909061].
  %%CITATION = GR-QC 9909061;%%
}

\lref\attract{ S.~Ferrara, R.~Kallosh and A.~Strominger, ``N=2
extremal black holes'', Phys.\ Rev.\ D {\bf 52}, 5412 (1995),
[arXiv:hep-th/9508072];
  %%CITATION = HEP-TH 9508072;%%
 A.~Strominger,
 ``Macroscopic Entropy of $N=2$ Extremal Black Holes'',
 Phys.\ Lett.\ B {\bf 383}, 39 (1996),
[arXiv:hep-th/9602111];
  %%CITATION = HEP-TH 9602111;%%
S.~Ferrara and R.~Kallosh, ``Supersymmetry and Attractors'', Phys.\
Rev.\ D {\bf 54}, 1514 (1996), [arXiv:hep-th/9602136];
  %%CITATION = HEP-TH 9602136;%%
``Universality of Supersymmetric Attractors'', Phys.\ Rev.\ D {\bf
54}, 1525 (1996), [arXiv:hep-th/9603090];
 %%CITATION = HEP-TH 9603090;%%
R.~Kallosh, A.~Rajaraman and W.~K.~Wong, ``Supersymmetric rotating
black holes and attractors'', Phys.\ Rev.\ D {\bf 55}, 3246 (1997),
[arXiv:hep-th/9611094];
  %%CITATION = HEP-TH 9611094;%%
A~Chou, R.~Kallosh, J.~Rahmfeld, S.~J.~Rey, M.~Shmakova and
W.~K.~Wong, ``Critical points and phase transitions in 5d
compactifications of M-theory''. Nucl.\ Phys.\ B {\bf 508}, 147
(1997); [arXiv:hep-th/9704142].
 %%CITATION = HEP-TH 9704142;%%
}

\lref\moore{G.~W.~Moore,``Attractors and arithmetic'',
[arXiv:hep-th/9807056];
%%CITATION = HEP-TH 9807056;%%
``Arithmetic and attractors'', [arXiv:hep-th/9807087];
  %%CITATION = HEP-TH 9807087;%%
``Les Houches lectures on strings and arithmetic'',
[arXiv:hep-th/0401049];
  %%CITATION = HEP-TH 0401049;%%
B.~R.~Greene and C.~I.~Lazaroiu, ``Collapsing D-branes in Calabi-Yau
moduli space. I'', Nucl.\ Phys.\ B {\bf 604}, 181 (2001),
[arXiv:hep-th/0001025]. }

%\ChamseddinePI
\lref\ChamseddinePI{
  A.~H.~Chamseddine, S.~Ferrara, G.~W.~Gibbons and R.~Kallosh,
  ``Enhancement of supersymmetry near 5d black hole horizon,''
  Phys.\ Rev.\ D {\bf 55}, 3647 (1997)
  [arXiv:hep-th/9610155].
  %%CITATION = HEP-TH 9610155;%%
}

\lref\denef{  %%CITATION = HEP-TH 0001025;%%
F.~Denef,``Supergravity flows and D-brane stability'', JHEP {\bf
0008}, 050 (2000), [arXiv:hep-th/0005049];
%%CITATION = HEP-TH 0005049;%%
``On the correspondence between D-branes and stationary supergravity
 solutions of type II Calabi-Yau compactifications'',
[arXiv:hep-th/0010222];
 %%CITATION = HEP-TH 0010222;%%
``(Dis)assembling special Lagrangians'', [arXiv:hep-th/0107152].
%%CITATION = HEP-TH 0107152;%%
  B.~Bates and F.~Denef,
   ``Exact solutions for supersymmetric stationary black hole composites,''
  [arXiv:hep-th/0304094].
}

\lref\OSV{H.~Ooguri, A.~Strominger and C.~Vafa, ``Black hole
attractors and the topological string'', Phys.\ Rev.\ D {\bf 70},
106007 (2004), [arXiv:hep-th/0405146];
 %%CITATION = HEP-TH 0405146;%%
}

\lref\moreOSV{ J.~de Boer, M.~C.~N.~Cheng, R.~Dijkgraaf, J.~Manschot
and E.~Verlinde, ``A farey tail for attractor black holes,'' JHEP
{\bf 0611}, 024 (2006) [arXiv:hep-th/0608059].
%%CITATION = JHEPA,0611,024;%%
}

%\DabholkarYR
\lref\DabholkarYR{
  A.~Dabholkar,
  ``Exact counting of black hole microstates,''
  Phys.\ Rev.\ Lett.\  {\bf 94}, 241301 (2005)
  [arXiv:hep-th/0409148].
  %%CITATION = PRLTA,94,241301;%%
}

 \lref\DDMP{
A.~Dabholkar, F.~Denef, G.~W.~Moore and B.~Pioline, ``Exact and
asymptotic degeneracies of small black holes'',
[arXiv:hep-th/0502157];  ``Precision counting of small black
holes,''
  JHEP {\bf 0510}, 096 (2005)
  [arXiv:hep-th/0507014].
  %%CITATION = JHEPA,0510,096;%%
  %%CITATION = HEP-TH 0502157;%%
}

\lref\curvcorr{A.~Dabholkar, ``Exact counting of black hole
microstates", [arXiv:hep-th/0409148],
%%CITATION = HEP-TH 0409148;%%
A.~Dabholkar, R.~Kallosh and A.~Maloney, ``A stringy cloak for a
classical singularity'', JHEP {\bf 0412}, 059 (2004),
[arXiv:hep-th/0410076].
 %%CITATION = HEP-TH 0410076;%%
} \lref\bkmicro{
 I.~Bena and P.~Kraus,
 ``Microscopic description of black rings in AdS/CFT'',
JHEP {\bf 0412}, 070 (2004)
  [arXiv:hep-th/0408186].
%%CITATION = HEP-TH 0408186;%%
} \lref\cgms{ M.~Cyrier, M.~Guica, D.~Mateos and A.~Strominger,
``Microscopic entropy of the black ring'', [arXiv:hep-th/0411187].
  %%CITATION = HEP-TH 0411187;%%
}

%\CardosoFP
\lref\CardosoFP{
  K.~Behrndt, G.~Lopes Cardoso, B.~de Wit, D.~Lust, T.~Mohaupt and
  W.~A.~Sabra,
  ``Higher-order black-hole solutions in N = 2 supergravity and
  Calabi-Yau
  string backgrounds,''
  Phys.\ Lett.\ B {\bf 429}, 289 (1998)
  [arXiv:hep-th/9801081];G.~Lopes Cardoso, B.~de Wit, D.~Lust,
  T.~Mohaupt,
  ``Corrections to macroscopic supersymmetric black-hole entropy'',
  Phys.\ Lett.\ B {\bf 451}, 309 (1999)
  [arXiv:hep-th/9812082].
  %%CITATION = HEP-TH 9812082;%%
  %%CITATION = HEP-TH 9801081;%%
    ``Macroscopic entropy formulae and non-holomorphic corrections for
  supersymmetric black holes'',
  Nucl.\ Phys.\ B {\bf 567}, 87 (2000)
  [arXiv:hep-th/9906094];
%G.~L.~Cardoso, B.~de Wit, J.~Kappeli and T.~Mohaupt, ``Examples of
%stationary BPS solutions in N = 2 supergravity theories  with
%$R^2$-interactions,'' Fortsch.\ Phys.\  {\bf 49}, 557 (2001)
%[arXiv:hep-th/0012232];
G.~Lopes Cardoso, B.~de Wit, J.~Kappeli , T.~Mohaupt
 ``Stationary BPS solutions in N = 2
supergravity with $R^2 $ interactions'', JHEP {\bf 0012}, 019 (2000)
[arXiv:hep-th/0009234];
%  ``Supersymmetric black hole solutions with $R^2$ interactions'',
%[arXiv:hep-th/0003157];
%  G.~Lopes Cardoso, B.~de Wit and T.~Mohaupt,
%  ``Area law corrections from state counting and supergravity'',
%  Class.\ Quant.\ Grav.\  {\bf 17}, 1007 (2000)
%  [arXiv:hep-th/9910179];
  %``Deviations from the area law for supersymmetric black holes'',
  %Fortsch.\ Phys.\  {\bf 48}, 49 (2000)
  %[arXiv:hep-th/9904005];
  }

\lref\hensken{  M.~Henningson and K.~Skenderis,
  ``The holographic Weyl anomaly,''
  JHEP {\bf 9807}, 023 (1998)
  [arXiv:hep-th/9806087].
  %%CITATION = HEP-TH 9806087;%%
  }

\lref\balkraus{  V.~Balasubramanian and P.~Kraus,
  ``A stress tensor for anti-de Sitter gravity,''
  Commun.\ Math.\ Phys.\  {\bf 208}, 413 (1999)
  [arXiv:hep-th/9902121].
  %%CITATION = HEP-TH 9902121;%%
  }

  %\HanakiPJ
\lref\HanakiPJ{
  K.~Hanaki, K.~Ohashi and Y.~Tachikawa,
  ``Supersymmetric completion of an R**2 term in five-dimensional
  supergravity,''
  [arXiv:hep-th/0611329].
  %%CITATION = HEP-TH 0611329;%%
}

%\KrausVZ
\lref\us{
  P.~Kraus and F.~Larsen,
  ``Microscopic black hole entropy in theories with higher derivatives,''
  JHEP {\bf 0509}, 034 (2005)
  [arXiv:hep-th/0506176].
  %%CITATION = JHEPA,0509,034;%%;
}

%\SahooVZ
\lref\SahooVZ{
  B.~Sahoo and A.~Sen,
  ``BTZ black hole with Chern-Simons and higher derivative terms,''
  JHEP {\bf 0607}, 008 (2006)
  [arXiv:hep-th/0601228].
  %%CITATION = JHEPA,0607,008;%%
}

%\GutowskiYV
\lref\GutowskiYV{
  J.~B.~Gutowski and H.~S.~Reall,
  ``General supersymmetric AdS(5) black holes,''
  JHEP {\bf 0404}, 048 (2004)
  [arXiv:hep-th/0401129].
  %%CITATION = HEP-TH 0401129;%%
}

%\LarsenXM
\lref\LarsenXM{
  F.~Larsen,
  ``The attractor mechanism in five dimensions,''
  [arXiv:hep-th/0608191].
  %%CITATION = HEP-TH 0608191;%%
}

\lref\MaldacenaBW{
  J.~M.~Maldacena and A.~Strominger,
  ``AdS(3) black holes and a stringy exclusion principle,''
  JHEP {\bf 9812}, 005 (1998)
  [arXiv:hep-th/9804085].
  %%CITATION = HEP-TH 9804085;%%
}

%\BanadosGQ
\lref\BanadosGQ{
  M.~Banados, M.~Henneaux, C.~Teitelboim and J.~Zanelli,
  ``Geometry of the (2+1) black hole,''
  Phys.Rev.D {\bf 66}, 010001 (2002)
  [arXiv:gr-qc/9302012].
  %%CITATION = GR-QC 9302012;%%
}

%\StromingerYG
\lref\StromingerYG{
  A.~Strominger,
  ``AdS(2) quantum gravity and string theory,''
  JHEP {\bf 9901}, 007 (1999)
  [arXiv:hep-th/9809027].
  %%CITATION = HEP-TH 9809027;%%
}

%\MohauptMJ
\lref\MohauptMJ{
  T.~Mohaupt,
  ``Black hole entropy, special geometry and strings,''
  Phys.Rev.D {\bf 66}, 010001 (2002)
  arXiv:hep-th/0007195;
  %%CITATION = HEP-TH 0007195;%%
  ``Supersymmetric black holes in string theory,''
  [arXiv:hep-th/0703035].
}

%\BehrndtHE
\lref\BehrndtHE{
  K.~Behrndt, G.~Lopes Cardoso and S.~Mahapatra,
  ``Exploring the relation between 4D and 5D BPS solutions,''
  Nucl.\ Phys.\  B {\bf 732}, 200 (2006)
  [arXiv:hep-th/0506251].
  %%CITATION = NUPHA,B732,200;%%
}

%\BenaNI
\lref\BenaNI{
  I.~Bena, P.~Kraus and N.~P.~Warner,
  ``Black rings in Taub-NUT,''
  Phys.\ Rev.\  D {\bf 72}, 084019 (2005)
  [arXiv:hep-th/0504142].
  %%CITATION = PHRVA,D72,084019;%%
}

%\GaiottoGF
\lref\GaiottoGF{
  D.~Gaiotto, A.~Strominger and X.~Yin,
  ``New connections between 4D and 5D black holes,''
  JHEP {\bf 0602}, 024 (2006)
  [arXiv:hep-th/0503217].
  %%CITATION = JHEPA,0602,024;%%
}

%\GaiottoXT
\lref\GaiottoXT{
  D.~Gaiotto, A.~Strominger and X.~Yin,
  ``5D black rings and 4D black holes,''
  JHEP {\bf 0602}, 023 (2006)
  [arXiv:hep-th/0504126].
  %%CITATION = JHEPA,0602,023;%%
}

%\GuicaIG
\lref\GuicaIG{
  M.~Guica, L.~Huang, W.~Li and A.~Strominger,
  ``R**2 corrections for 5D black holes and rings,''
  JHEP {\bf 0610}, 036 (2006)
  [arXiv:hep-th/0505188].
  %%CITATION = JHEPA,0610,036;%%
}

%\KrausZM
\lref\KrausZM{
  P.~Kraus and F.~Larsen,
  ``Holographic gravitational anomalies,''
  JHEP {\bf 0601}, 022 (2006)
  [arXiv:hep-th/0508218].
  %%CITATION = HEP-TH 0508218;%%
}

%\GaiottoNS
\lref\GaiottoNS{
  D.~Gaiotto, A.~Strominger and X.~Yin,
  ``From AdS(3)/CFT(2) to black holes / topological strings,''
  [arXiv:hep-th/0602046].
  %%CITATION = HEP-TH/0602046;%%
}

%\MohauptMJ
\lref\MohauptMJ{
  T.~Mohaupt,
  ``Black hole entropy, special geometry and strings,''
  Fortsch.\ Phys.\  {\bf 49}, 3 (2001)
  [arXiv:hep-th/0007195].
  %%CITATION = FPYKA,49,3;%%
}

%\PiolineNI
\lref\PiolineNI{
  B.~Pioline,
  ``Lectures on black holes, topological strings and quantum attractors,''
  Class.\ Quant.\ Grav.\  {\bf 23}, S981 (2006)
  [arXiv:hep-th/0607227].
  %%CITATION = CQGRD,23,S981;%%
}

%\KrausWN
\lref\KrausWN{
  P.~Kraus,
  ``Lectures on black holes and the AdS(3)/CFT(2) correspondence,''
  [arXiv:hep-th/0609074].
  %%CITATION = HEP-TH/0609074;%%
}

%\DabholkarTB
\lref\DabholkarTB{
  A.~Dabholkar, A.~Sen and S.~P.~Trivedi,
  ``Black hole microstates and attractor without supersymmetry,''
  JHEP {\bf 0701}, 096 (2007)
  [arXiv:hep-th/0611143].
  %%CITATION = JHEPA,0701,096;%%
}

%\BenaAE
\lref\rings{
  I.~Bena and P.~Kraus,
  ``R**2 corrections to black ring entropy,''
  arXiv:hep-th/0506015;
  %%CITATION = HEP-TH/0506015;%%
  N.~Iizuka and M.~Shigemori,
  ``A note on D1-D5-J system and 5D small black ring,''
  JHEP {\bf 0508}, 100 (2005)
  [arXiv:hep-th/0506215]; A.~Dabholkar, N.~Iizuka, A.~Iqubal and M.~Shigemori,
  ``Precision microstate counting of small black rings,''
  Phys.\ Rev.\ Lett.\  {\bf 96}, 071601 (2006)
  [arXiv:hep-th/0511120]; A.~Dabholkar, N.~Iizuka, A.~Iqubal, A.~Sen and M.~Shigemori,
  ``Spinning strings as small black rings,''
 [arXiv:hep-th/0611166].
}

%\TachikawaSZ
\lref\TachikawaSZ{
  Y.~Tachikawa,
  ``Black hole entropy in the presence of Chern-Simons terms,''
  Class.\ Quant.\ Grav.\  {\bf 24}, 737 (2007)
  [arXiv:hep-th/0611141].
  %%CITATION = CQGRD,24,737;%%
}

%\KugoHN
\lref\KugoHN{
  T.~Kugo and K.~Ohashi,
  ``Supergravity tensor calculus in 5D from 6D,''
  Prog.\ Theor.\ Phys.\  {\bf 104}, 835 (2000)
  [arXiv:hep-ph/0006231]; T.~Fujita and K.~Ohashi,
  ``Superconformal tensor calculus in five dimensions,''
  Prog.\ Theor.\ Phys.\  {\bf 106}, 221 (2001)
  [arXiv:hep-th/0104130].
  %%CITATION = PTPKA,104,835;%%
}

%\BergshoeffHC
\lref\BergshoeffHC{
  E.~Bergshoeff, S.~Cucu, M.~Derix, T.~de Wit, R.~Halbersma and A.~Van Proeyen,
  ``Weyl multiplets of N = 2 conformal supergravity in five dimensions,''
  JHEP {\bf 0106}, 051 (2001)
  [arXiv:hep-th/0104113].
  %%CITATION = JHEPA,0106,051;%%
  ``N = 2 supergravity in five dimensions revisited,''
  Class.\ Quant.\ Grav.\  {\bf 21}, 3015 (2004)
  [Class.\ Quant.\ Grav.\  {\bf 23}, 7149 (2006)]
  [arXiv:hep-th/0403045].
  %%CITATION = CQGRD,23,7149;%%
}

%\DabholkarDQ
\lref\DabholkarDQ{
  A.~Dabholkar, R.~Kallosh and A.~Maloney,
  ``A stringy cloak for a classical singularity,''
  JHEP {\bf 0412}, 059 (2004)
  [arXiv:hep-th/0410076];
  %%CITATION = JHEPA,0412,059;%%
%}
%\HubenyJI
%\lref\HubenyJI{
  V.~Hubeny, A.~Maloney and M.~Rangamani,
  ``String-corrected black holes,''
  JHEP {\bf 0505}, 035 (2005)
  [arXiv:hep-th/0411272].
  %%CITATION = JHEPA,0505,035;%%
}

\lref\senstretch{A.~Sen,
  ``How does a fundamental string stretch its horizon?,''
  JHEP {\bf 0505}, 059 (2005)
  [arXiv:hep-th/0411255];
  %%CITATION = HEP-TH 0411255;%%
  }

%\KrausNB
\lref\KrausNB{
  P.~Kraus and F.~Larsen,
  ``Partition functions and elliptic genera from supergravity,''
  JHEP {\bf 0701}, 002 (2007)
  [arXiv:hep-th/0607138].
  %%CITATION = JHEPA,0701,002;%%
}

%\CastroSD
\lref\CastroSD{
  A.~Castro, J.~L.~Davis, P.~Kraus and F.~Larsen,
  ``5D attractors with higher derivatives,''
  [arXiv:hep-th/0702072].
  %%CITATION = HEP-TH/0702072;%%
}

%\CastroHC
\lref\CastroHC{
  A.~Castro, J.~L.~Davis, P.~Kraus and F.~Larsen,
  ``5D Black Holes and Strings with Higher Derivatives,''
  [arXiv:hep-th/0703087].
  %%CITATION = HEP-TH/0703087;%%
}

\lref\moreCDKL{
  A.~Castro, J.~L.~Davis, P.~Kraus and F.~Larsen.
  In progress.
}

%\CadavidBK
\lref\CadavidBK{
  A.~C.~Cadavid, A.~Ceresole, R.~D'Auria and S.~Ferrara,
  %``Eleven-dimensional supergravity compactified on Calabi-Yau threefolds,''
  Phys.\ Lett.\  B {\bf 357}, 76 (1995)
  [arXiv:hep-th/9506144].
  %%CITATION = PHLTA,B357,76;%%
}

%\ShmakovaNZ
\lref\ShmakovaNZ{
  M.~Shmakova,
  ``Calabi-Yau black holes,''
  Phys.\ Rev.\  D {\bf 56}, 540 (1997)
  [arXiv:hep-th/9612076].
  %%CITATION = PHRVA,D56,540;%%
}

%\ChouBA
\lref\ChouBA{
  A.~Chou, R.~Kallosh, J.~Rahmfeld, S.~J.~Rey, M.~Shmakova and W.~K.~Wong,
  ``Critical points and phase transitions in 5d compactifications of
  M-theory,''
  Nucl.\ Phys.\  B {\bf 508}, 147 (1997)
  [arXiv:hep-th/9704142].
  %%CITATION = NUPHA,B508,147;%%
}

%\MaldacenaDE
\lref\MaldacenaDE{
  J.~M.~Maldacena, A.~Strominger and E.~Witten,
  ``Black hole entropy in M-theory,''
  JHEP {\bf 9712}, 002 (1997)
  [arXiv:hep-th/9711053].
  %%CITATION = JHEPA,9712,002;%%
}

%\VafaGR
\lref\VafaGR{
  C.~Vafa,
  ``Black holes and Calabi-Yau threefolds,''
  Adv.\ Theor.\ Math.\ Phys.\  {\bf 2}, 207 (1998)
  [arXiv:hep-th/9711067].
  %%CITATION = 00203,2,207;%%
}

%\HarveyBX
\lref\HarveyBX{
  J.~A.~Harvey, R.~Minasian and G.~W.~Moore,
  ``Non-abelian tensor-multiplet anomalies,''
  JHEP {\bf 9809}, 004 (1998)
  [arXiv:hep-th/9808060].
  %%CITATION = JHEPA,9809,004;%%
}

%\BenaAY
\lref\BenaAY{
  I.~Bena and P.~Kraus,
  `Microstates of the D1-D5-KK system,''
  Phys.\ Rev.\  D {\bf 72}, 025007 (2005)
  [arXiv:hep-th/0503053]; ``Microscopic description of black rings in AdS/CFT,''
  JHEP {\bf 0412}, 070 (2004)
  [arXiv:hep-th/0408186].
  %%CITATION = PHRVA,D72,025007;%%
}

%\ElvangSA
\lref\ElvangSA{
  H.~Elvang, R.~Emparan, D.~Mateos and H.~S.~Reall,
  ``Supersymmetric 4D rotating black holes from 5D black rings,''
  JHEP {\bf 0508}, 042 (2005)
  [arXiv:hep-th/0504125].
  %%CITATION = JHEPA,0508,042;%%
}

%\StromingerSH
\lref\StromingerSH{
  A.~Strominger and C.~Vafa,
  ``Microscopic Origin of the Bekenstein-Hawking Entropy,''
  Phys.\ Lett.\  B {\bf 379}, 99 (1996)
  [arXiv:hep-th/9601029].
  %%CITATION = PHLTA,B379,99;%%
}

%\BreckenridgeIS
\lref\BreckenridgeIS{
  J.~C.~Breckenridge, R.~C.~Myers, A.~W.~Peet and C.~Vafa,
  ``D-branes and spinning black holes,''
  Phys.\ Lett.\  B {\bf 391}, 93 (1997)
  [arXiv:hep-th/9602065].
  %%CITATION = PHLTA,B391,93;%%
}

%\GauntlettNW
\lref\GauntlettNW{
  J.~P.~Gauntlett, J.~B.~Gutowski, C.~M.~Hull, S.~Pakis and H.~S.~Reall,
  ``All supersymmetric solutions of minimal supergravity in five dimensions,''
  Class.\ Quant.\ Grav.\  {\bf 20}, 4587 (2003)
  [arXiv:hep-th/0209114].
  %%CITATION = CQGRD,20,4587;%%
}

%\GoldsteinKM
\lref\GoldsteinKM{
  K.~Goldstein and R.~P.~Jena,
  ``One entropy function to rule them all,''
  [arXiv:hep-th/0701221].
  %%CITATION = HEP-TH/0701221;%%
}

%\GuicaWD
\lref\GuicaWD{
  M.~Guica and A.~Strominger,
  ``Cargese lectures on string theory with eight supercharges,''
  arXiv:0704.3295 [hep-th].
  %%CITATION = ARXIV:0704.3295;%%
}

%\AstefaneseiDD
\lref\AstefaneseiDD{
  D.~Astefanesei, K.~Goldstein, R.~P.~Jena, A.~Sen and S.~P.~Trivedi,
  ``Rotating attractors,''
  JHEP {\bf 0610}, 058 (2006)
  [arXiv:hep-th/0606244].
  %%CITATION = JHEPA,0610,058;%%
}

%\MarolfCB
\lref\MarolfCB{
  D.~Marolf,
  ``Chern-Simons terms and the three notions of charge,''
  [arXiv:hep-th/0006117].
  %%CITATION = HEP-TH/0006117;%%
}
%\HanakiMB
\lref\HanakiMB{
  K.~Hanaki, K.~Ohashi and Y.~Tachikawa,
  ``Comments on Charges and Near-Horizon Data of Black Rings,''
  arXiv:0704.1819 [hep-th].
  %%CITATION = ARXIV:0704.1819;%%
}

%\BreckenridgeIS
\lref\BreckenridgeIS{
  J.~C.~Breckenridge, R.~C.~Myers, A.~W.~Peet and C.~Vafa,
  ``D-branes and spinning black holes,''
  Phys.\ Lett.\  B {\bf 391}, 93 (1997)
  [arXiv:hep-th/9602065].
  %%CITATION = PHLTA,B391,93;%%
}

%\SuryanarayanaRK
\lref\SuryanarayanaRK{
  N.~V.~Suryanarayana and M.~C.~Wapler,
  ``Charges from Attractors,''
  arXiv:0704.0955 [hep-th].
  %%CITATION = ARXIV:0704.0955;%%
}

%\AstefaneseiDD
\lref\AstefaneseiDD{
  D.~Astefanesei, K.~Goldstein, R.~P.~Jena, A.~Sen and S.~P.~Trivedi,
  ``Rotating attractors,''
  JHEP {\bf 0610}, 058 (2006)
  [arXiv:hep-th/0606244].
  %%CITATION = JHEPA,0610,058;%%
}

%\CardosoRG
\lref\CardosoRG{
  G.~L.~Cardoso, J.~M.~Oberreuter and J.~Perz,
  ``Entropy function for rotating extremal black holes in very special
  geometry,''
  [arXiv:hep-th/0701176].
  %%CITATION = HEP-TH/0701176;%%
}

%\MoralesGM
\lref\MoralesGM{
  J.~F.~Morales and H.~Samtleben,
  ``Entropy function and attractors for AdS black holes,''
  JHEP {\bf 0610}, 074 (2006)
  [arXiv:hep-th/0608044].
  %%CITATION = JHEPA,0610,074;%%
}

%\CaiUW
\lref\CaiUW{
  R.~G.~Cai and D.~W.~Pang,
  ``On Entropy Function for Supersymmetric Black Rings,''
  JHEP {\bf 0704}, 027 (2007)
  [arXiv:hep-th/0702040].
  %%CITATION = JHEPA,0704,027;%%
}

%\DabholkarZA
\lref\DabholkarZA{
  A.~Dabholkar, N.~Iizuka, A.~Iqubal, A.~Sen and M.~Shigemori,
  ``Spinning strings as small black rings,''
  JHEP {\bf 0704}, 017 (2007)
  [arXiv:hep-th/0611166].
  %%CITATION = JHEPA,0704,017;%%
}

%\AlishahihaNN
\lref\AlishahihaNN{
 M.~Alishahiha,
 ``On $R^2$ corrections for 5D black holes,''
 [arXiv:hep-th/0703099].
 %%CITATION = HEP-TH/0703099;%%
}

%\ElvangSA
\lref\ElvangSA{
  H.~Elvang, R.~Emparan, D.~Mateos and H.~S.~Reall,
  ``Supersymmetric 4D rotating black holes from 5D black rings,''
  JHEP {\bf 0508}, 042 (2005)
  [arXiv:hep-th/0504125].
  %%CITATION = JHEPA,0508,042;%%
}

%\KutasovZH
\lref\KutasovZH{
  D.~Kutasov, F.~Larsen and R.~G.~Leigh,
  ``String theory in magnetic monopole backgrounds,''
  Nucl.\ Phys.\  B {\bf 550}, 183 (1999)
  [arXiv:hep-th/9812027].
  %%CITATION = NUPHA,B550,183;%%
}

%\VafaGR
\lref\VafaGR{
  C.~Vafa,
  ``Black holes and Calabi-Yau threefolds,''
  Adv.\ Theor.\ Math.\ Phys.\  {\bf 2}, 207 (1998)
  [arXiv:hep-th/9711067].
  %%CITATION = 00203,2,207;%%
}

%\HuangSB
\lref\HuangSB{
  M.~x.~Huang, A.~Klemm, M.~Marino and A.~Tavanfar,
  ``Black Holes and Large Order Quantum Geometry,''
  arXiv:0704.2440 [hep-th].
  %%CITATION = ARXIV:0704.2440;%%
}

%%%%%%%%%%%%%%%%%%%%%%%%%%%%%%%%%%%%%%%%%%%%%%
%%%%%%%%%%%%%%%%%%%%%%%%%%%
% Title
%%%%%%%%%%%%%%%%%%%%%%%%%%%%%%%%%%%%%%%%%%%%%%
%%%%%%%%%%%%%%%%%%%%%%%%%%%

\Title{\vbox{\baselineskip12pt
%\hbox{hep-th/0508218}
%\hbox{UCLA-05-TEP-XX} \hbox{MCTP-XX-XX}
}} {\vbox{\centerline {Precision Entropy of Spinning Black Holes}}}
%\medskip\vbox{\centerline {more title}}} }
% Stringy Black Holes with Spin
\centerline{Alejandra
Castro$^\dagger$\foot{aycastro@umich.edu}, Joshua L.
Davis$^{\spadesuit}$\foot{davis@physics.ucla.edu}, Per
Kraus$^{\spadesuit}$\foot{pkraus@ucla.edu}, and Finn
Larsen$^\dagger$\foot{larsenf@umich.edu}}
\bigskip
\centerline{${}^\dagger$\it{Department of Physics
and Michigan Center for Theoretical Physics,
}} \centerline{\it{University of Michigan, Ann
Arbor, MI 48109-1120, USA.}}\vskip.2cm
\centerline{${}^{\spadesuit}$\it{Department of Physics and
Astronomy, UCLA,}}\centerline{\it{ Los Angeles, CA 90095-1547,
USA.}}

\baselineskip15pt

\vskip .3in

\centerline{\bf Abstract} We construct spinning black hole
solutions in five dimensions that take into account the mixed
gauge-gravitational Chern-Simons term and its supersymmetric
completion. The resulting  entropy formula is discussed from
several points of view.  We  include a Taub-NUT base space in
order to test recent conjectures relating  5D black holes to 4D
black holes and the topological string.  Our explicit results show
that certain charge shifts have to be taken into account for these
relations to hold.  We also compute  corrections to the entropy of
black rings in terms of near horizon data.

%We find the relation between horizon charges and asymptotic
%charges and also exhibit the relation between 4D and 5D
%descriptions by introducing a Taub-NUT base space. Our work gives
%a detailed picture of the entropy and the geometry which corrects
%earlier arguments motivated by topological string theory.

%%%
\Date{May, 2007}
%%%%%%%%%%%%%%%%%%%%%%%%%%%%%%%%%%%%%%%%%%%%%%
%%%%%%%%%%%%%%%%%%%%%%%%%%%
% Main text begins here
%%%%%%%%%%%%%%%%%%%%%%%%%%%%%%%%%%%%%%%%%%%%%%
%%%%%%%%%%%%%%%%%%%%%%%%%%%
\baselineskip14pt

\newsec{Introduction}

The importance of black holes for quantum gravity and string theory
has motivated a sustained effort to achieve a computational control
of black hole entropy that goes beyond the leading
Bekenstein-Hawking area law
\refs{\MaldacenaDE,\CardosoFP,\OSV,\senrescaled,\DabholkarYR,\DDMP,\us,\SenWA}.
While much has been achieved, there are still many unanswered
questions  (for recent reviews summarizing the current state of the
subject, see \refs{\MohauptMJ,\PiolineNI,\KrausWN,\GuicaWD}). On the
gravity side, the leading corrections to the entropy come from
higher derivative terms in the spacetime effective action, and we
would certainly like to know how these affect the standard black
hole solutions of string theory.   However, the results in this
direction have so far been limited to 4D black holes, which is
surprising given that the simplest supersymmetric black holes in
string theory arise in 5D.   We have recently begun to fill this gap
\refs{\CastroSD,\CastroHC} (see also \AlishahihaNN).  In this paper
we continue this program by constructing asymptotically flat
spinning black holes with higher derivative corrections taken into
account. Our solutions are generalizations of the BMPV solution \BreckenridgeIS.
They are  simple enough
that we can be quite explicit, yet intricate enough that we can
shed light on a number of important conceptual issues.

The setting for our analysis is 5D supergravity corrected by the
mixed gauge-gravitational Chern-Simons term
\eqn\aaa{ {c_{2I}\over 24\cdot 16\pi^2} \int A^I\wedge {\rm Tr}
R^2~, }
and terms related to this by supersymmetry. We use the off-shell
formalism which has supersymmetry transformations that do not
depend on the explicit Lagrangian. The supersymmetric completion
of \aaa\ was constructed in this formalism in \HanakiPJ. Taking
advantage of the universal supersymmetry variations, and also
using the complete action, we find the  solution for the spinning
black hole.

The next step is to determine the Bekenstein-Hawking-Wald \wald\
entropy of the black hole. The near horizon geometry consists of a
circle fibered over AdS$_2\times S^2$.  After KK reduction on the
circle, Wald's  entropy formula is equivalent to entropy
extremization
\refs{\SenWA,\SahooVZ,\AstefaneseiDD,\MoralesGM,\CardosoRG,\GoldsteinKM}.
A well known  subtlety in this procedure arises from the presence
of Chern-Simons terms, since these are not  gauge invariant
\SahooVZ. After this is taken into account,  we find the entropy
of a spinning black hole with higher derivative corrections.

Our result for the entropy is simplest when  expressed in terms of
the near horizon moduli.    In this form we can also demonstrate
precise agreement with results inferred from 4D black holes, the
topological string, and the 4D/5D connection
\refs{\MohauptMJ,\OSV,\GuicaWD,\GuicaIG}.   However,  the more
physically relevant result is the entropy expressed in terms of
the conserved charges,  and in these variables the relation with
the 4D results exhibits some new features.

The 5D electric charges are defined unambiguously in terms of flux
integrals over a $S^3$ at infinity surrounding the black hole. The
comparison to 4D black holes is made by placing the 5D black hole
at the tip of a Taub-NUT space.   Taub-NUT is asymptotically
$\IR^3\times S^1$, and a 4D black hole is obtained via KK
reduction on the $S^1$.   The 4D electric charges are thereby
defined via  flux integrals over an asymptotic $S^2$.  We can
think of recovering the 5D black hole by sending the radius of the
circle (which is a modulus) to infinity
\refs{\GaiottoGF,\ElvangSA,\GaiottoXT,\BenaNI,\BehrndtHE}.

At lowest (i.e. two-derivative) order, the 4D and 5D electric
charges are equal, and in the literature it seems to be assumed
that this holds in general.  However, we show explicitly that the
charges are different in the presence of higher derivatives.  In
particular, the electric charges differ by $\Delta q_I = {1 \over
24}c_{2I}$.  The reason is simple: the operations of computing the
flux integrals and decompactifying the Taub-NUT circle do not
commute.  This in turn follows from the fact that the Taub-NUT
space itself carries a delocalized electric charge proportional to
its Euler number, as implied by the Chern-Simons term \aaa.

Angular momentum adds further structure, and we find that another
higher derivative shift is required to relate $J$ to the
corresponding 4D electric charge $q_0$.  Our conclusion is that
all these shifts need to be taken into account in order to use the
4D/5D connection to reproduce the correct 5D entropy formulas
derived here.

While the main topic of this paper is 5D black holes, our entropy
analysis can be easily extended to the case of black rings.  We
thereby find the corrected black ring entropy formula, albeit
expressed in terms of near horizon data. Giving an expression in
terms of the charges of the ring requires knowledge of the full
asymptotically flat solution, which is not yet available
(alternatively, one might try to employ the techniques developed
in \refs{\SuryanarayanaRK,\HanakiMB}).

This paper is organized as follows. In section 2 we outline the
derivation of the spinning black hole solution and discuss some of
its properties. Some further details are provided in the Appendix.
In section 3 we derive the black hole entropy using entropy
extremization.  As an aside, we also find the entropy of the black
ring with higher derivatives. In section 4 we discuss
interpretational issues with emphasis on aspects related to the
definition of charge. We explain why results motivated by 4D
topological string theory fail to capture the full story. Finally,
we construct the spinning black hole on a Taub-NUT base space and
use this to  carry out the 4D-5D reduction explicitly.

\newsec{5D spinning black hole solutions}
We want to find the rotating supersymmetric  black holes in five
dimensions with higher derivatives taken into account. The procedure
for deriving the solution is the same as in the spherically
symmetric case \CastroHC\ so we shall focus on results rather than
methodology. Some details of our derivation  are given in Appendix
A.

\subsec{The supersymmetry conditions} The starting point is an
{\it ansatz} for the solution. Since the supersymmetry variations
in the off-shell formalism are unaffected by the presence of
higher derivatives terms in the action,  the form of the solution
is the same as in the two-derivative context \GauntlettNW.  In
particular, supersymmetry implies the existence of a timelike
Killing vector, which we build in by writing
\eqn\aa{ds^2=e^{4U(x)}(dt+\omega)^2-e^{-2U(x)}h_{mn} dx^m dx^n~,}
where $\omega=\omega_i(x)dx^i$ is a one-form on the 4D base
manifold equipped with metric $h_{mn} dx^m dx^n$. The base space
is generally Hyper-K\"{a}hler; for the present it is just taken to
be flat space (we discuss the case of Taub-NUT later). We will use
the obvious local frame
\eqn\ab{e^{\hat{0}}=e^{2U}(dt+\omega)~,\quad
e^{\hat{i}}=e^{-U}dx^i~.}

The matter in the theory consists of  $n_V$ vector multiplets of
${\cal N} =2 $  supersymmetry. Supersymmetry relates the gauge field
strength in each multiplet to the  corresponding scalar field
through the attractor flow
\eqn\ac{F^I=d(M^Ie^{\hat{0}})~.
}
Generally, supersymmetry also permits the addition to $F^I$ of an
anti-self-dual form on the base space.   Such a contribution is
needed for black ring solutions, but vanishes for the black hole
solutions considered here.  With this restriction, we also have
that $d\omega$ is self-dual\foot{$\star_4$ denotes the dual taken
with respect to the metric $h_{mn}dx^m dx^n$.}
\eqn\aca{ \star_4 d\omega = d\omega~.}

Supersymmetry further determines the auxiliary fields completely
in terms of  the geometry \aa. The auxiliary two-form is fixed to
be
\eqn\ad{ v = - {3\over 4} d e^{\hat 0} = -{3\over 4} \left(
2\partial_i U e^{U} e^{\ih}e^{\hat 0} + {1\over 2} e^{2U}
d\omega_{\ih\jh} e^{\ih} e^{\jh} \right)~, }
and the auxiliary scalar is determined as
\eqn\ae{\eqalign{D&=3e^{2U}\left(\nabla^2U-6(\nabla
U)^2\right)+{3\over 2}e^{8U}(d\omega)^2~.}}

\subsec{Equations of motion}

At this point the  constraints of supersymmetry have been
exhausted and we must use the explicit action \HanakiPJ. First of
all, we need the equations of motion for the gauge field, namely
the Maxwell equation
\eqn\afa{2 \nabla_\mu \left( {\p \Lc \over \p F^{I}_{~\mu\nu}}
\right) = {\p \Lc \over \p A^I_\nu}  ~.}
It is straightforward in principle (although tedious in practice)
to insert a solution of the general form \aa-\ae\ into the Maxwell
equation \afa. After reorganization, we find that the spatial
components of the equation are satisfied automatically. We also
find that the temporal component can be cast in the simple form
\eqn\ah{\nabla^2\left[e^{-2U}M_I - {c_{2I}\over8}\left((\nabla
U)^2-{1\over 12}e^{6U}(d\omega)^2\right)\right]=0~,}
with
\eqn\aga{ M_I = {1 \over 2}c_{IJK} M^J M^K~.}
All indices in \ah\ are contracted with the base space metric
$h_{mn}$, e.g.,
\eqn\aia{ (d\omega)^2 = h^{mn}h^{pq}d\omega_{mp}d\omega_{nq}~.}
\ah\ is the generalized Gauss' law, and is simply a harmonic
equation on the flat base space.\foot{Later we will find that a
curved base metric induces a source on the right hand side of this
equation.} We will later discuss how conserved charges can be read
off from this equation, with nontrivial shifts due to higher
derivatives encoded in the term proportional to $c_{2I}$. At this
point we just note that the one-form $\omega$ enters Gauss' law
when higher derivatives are taken into account. The decoupling
between angular momentum and radial evolution found in the leading
order theory is therefore not preserved in general.

In order to fully specify the solution we also  need the equation
of motion for the auxiliary  field $D$. It is
\eqn\ai{{\cal N}=1-{c_{2I}\over72}(F^{I}_{\mu\nu}v^{\mu\nu}+M^ID)~.}
where ${\cal N} = {1\over 6} c_{IJK} M^I M^J M^K$.
Inserting \aa-\ae\ for the spinning black hole we find
\eqn\aj{ {1\over 6} c_{IJK} M^I M^J M^K=1-{c_{2I}\over24}
\left[e^{2U}M^I\left(\nabla^2U-4(\nabla U)^2+{1\over
4}e^{6U}(d\omega)^2\right)+e^{2U}\nabla^i M^I\nabla_iU\right]~.}
In the two-derivative theory the scalar fields are constrained by the special geometry
condition ${\cal N}=1$. In the corrected theory we must instead impose the much more
complicated condition \aj.

\subsec{Assembling the solution}

We have now determined all the necessary equations and it only
remains to solve them.   This is simplified by writing the flat
base space in the Gibbons-Hawking coordinates\foot{The
transformation $\rho = {r^2 \over 4},~ x^5 =
\tilde{\phi}+\tilde{\psi},~ \phi = \tilde{\phi}-\tilde{\psi},~
\theta = 2\tilde{\theta}$ brings the line element to the form
$ds^2 = dr^2 + r^2(d\tilde{\theta}^2 + \sin^2 \tilde{\theta}
d\tilde{\psi}^2 +\cos^2 \tilde{\theta}d\tilde{\phi}^2 )$.}
\eqn\ba{ h_{mn} dx^m dx^n = \rho (dx^5 + \cos\theta d\phi)^2 +
{1\over\rho}\left(d\rho^2 + \rho^2(d\theta^2+\sin^2\theta
d\phi^2)\right)~, }
with $x^5 \cong x^5 + 4\pi$.

Let us recall the BMPV solution to the two derivative theory
\BreckenridgeIS. When written in Gibbons-Hawking coordinates the
one-form on the base space takes the form\foot{We use units $G_5 =
{\pi\over 4}$. In these  units the angular momentum and the
charges are quantized as integers.}
\eqn\bb{ \omega =  {J\over 8\rho}(dx^5 + \cos\theta d\phi)~, }
so that
\eqn\bc{ d\omega = -{J\over 8\rho^2} ( e^{\hat\rho}e^{\hat 5} +
 e^{\hat\theta}e^{\hat\phi})~,
}
in the obvious orthonormal frame on the base space. In this form
the self-duality condition $d\omega=\star_4 d\omega$ is manifest.
In fact, the $\rho$-dependence of $d\omega$ is completely
determined by the Bianchi identity and the self-duality condition.
Therefore \bb-\bc\ will be maintained when higher derivatives are
taken into account.

Let us next turn to the generalized Gauss' law \ah. As  already
noted, this is just a harmonic equation. Writing out the Laplacian
in Gibbons-Hawking coordinates we are lead to introduce the
harmonic function
\eqn\bd{H_I = M_I^\infty + {q_I\over 4\rho}= e^{-2U}M_I -
{c_{2I}\over8}\left((\nabla U)^2-{1\over
12}e^{6U}(d\omega)^2\right)~,}
where the constants of integration $M_I^\infty$ are identified
with the  asymptotic moduli.

The solution we seek is specified by the conserved charges  $(J,
q_I)$ and the asymptotic moduli. With these inputs, the one-form
$\omega$ was given in \bb, the gauge field strengths were found in
\ac, and \bd\ determines the scalar fields as
\eqn\be{
M_I(\rho) = e^{2U}\left[ M_I^\infty + {q_I\over 4\rho}+
  {c_{2I}\over8}\left((\nabla
U)^2-{1\over 12}e^{6U}(d\omega)^2\right) \right] ~.}

Up to this point the solution has been given not only in terms of
the conserved charges, but also in terms of the metric function
$U(\rho)$, which has not yet been computed. This function is
determined by the constraint \aj. In order to make this additional
equation completely explicit we should first invert the equation
\aga\ that
 determines $M^I$ in terms of $M_I$. The result should
be inserted in \aj, which then becomes an  second order ordinary
differential equation that  can be   easily integrated numerically
to find $U(\rho)$. In \CastroHC\ we carried out this procedure for
some examples with spherical symmetry. The rotating solution is
qualitatively similar, but not identical. In particular, we
mention again that the radial profile depends on the angular
momentum when higher derivative corrections are taken into
account.

\subsec{Near horizon geometry} We are especially interested in the
near horizon region, and here we can make the geometry more
explicit. In order to do that we consider a radial function of the
form
\eqn\bg{
e^{2U} = {\rho\over\ell^2}~.
}
The parameter $\ell$ sets the physical scale of the solution. We
will see later that it can be identified with the radii of a near
horizon $AdS_2\times S^2$.\foot{In the nonrotating case there is a
near horizon $AdS_2\times S^3$ with radii
$\ell_A=\half\ell_S=\ell$. This was the notation used in
\CastroHC.} With this radial function the scalar fields \be\
reduce to the constants
\eqn\bh{ M_I = {1\over 4\ell^2}\left( q_I + {c_{2I}\over 8}\left(1 -
{1\over 48\ell^6}J^2\right)\right)~. }
These are the attractor values for the moduli in the geometry modified by higher
derivatives. In particular, the attractor values depend on the conserved charges
alone, and not the asymptotic moduli.

The constraint equation \aj\ also becomes an algebraic relation
\eqn\bi{
{1\over 6} c_{IJK} M^I M^J M^K
=   1 + {c_{2I} M^I \over 48\ell^2}\left( 1 - {J^2\over 32\ell^6}\right)~.
}
Taken together with the relations \aga\ we have a set of algebraic
equations that determine the near horizon geometry completely. In
order to solve these equations it is convenient to introduce the
scaled variables
\eqn\bj{\eqalign{
{\hat M}^I & = 2\ell M^I~,\cr
{\hat J} & = {1\over 8\ell^3}J~.
}}
We then have the following procedure: given asymptotic charges $(J,q_I)$
we find the rescaled variables $({\hat J}, {\hat M}^I)$ by solving the equations
\bh-\bi\ written in the form
\eqn\bk{\eqalign{
J & = \left( {1\over 3!} c_{IJK} {\hat M}^I {\hat M}^J {\hat M}^K
- {c_{2I}{\hat M}^I\over 12}( 1 - 2{\hat J}^2)\right){\hat J}~,\cr
q_I & = {1\over 2} c_{IJK} {\hat M}^J {\hat M}^K - {c_{2I}\over 8}\left( 1 - {4\over 3} {\hat J}^2\right)~.
}}
With the solution in hand we compute
\eqn\bl{\eqalign{
\ell^3 &= {1\over 8}\left( {1\over 3!} c_{IJK} {\hat M}^I {\hat M}^J {\hat M}^K
- {c_{2I}{\hat M}^I\over 12}( 1 - 2{\hat J}^2)\right)~,\cr
M^I & = {1\over 2\ell}{\hat M}^I~,
}}
to find the values for the physical scale of the solution $\ell$
and the physical moduli $M^I$, written as functions of $(J,q_I)$.

In general it is of course rather difficult to invert \bk\
explicitly. This is the situation also before higher derivative
corrections have been taken into account and/or if angular
momentum is neglected.

The formulae can be made more explicit for large charges. Let us
define the dual charges $q^I$ through \eqn\bla{ q_I = {1\over 2}
c_{IJK} q^J q^K~. } We also define  \eqn\blb{ Q^{3/2} = {1\over
3!} c_{IJK} q^I q^J q^K~, } and  \eqn\blc{ C_{IJ} = c_{IJK} q^K~.
} Each of these quantities depend on charges and Calabi-Yau data
but not on moduli.

With the definitions \bla-\blc\ we can invert \bk\ for large
charges (i.e. expand to first order in $c_{2I}$) and find
\eqn\bld{\eqalign{ {\hat M}^I &= q^I + {1\over 8} \left( 1 -
{4\over 3} {J^2\over Q^3}\right)C^{IJ}c_{2J} + \ldots~,\cr {\hat
J} & = {J\over Q^{3/2}} \left( 1 + {c_{2}\cdot q\over 48
Q^{3/2}}\left[1-4{J^2 \over Q^3}\right] \right)+\ldots~. }}
Then \bl\ gives the physical
scale of the geometry and the physical moduli as
\eqn\ble{\eqalign{ \ell & = {1\over 2}Q^{1/2}\left( 1  -
{c_{2}\cdot q\over 144 Q^{3/2}}\left[1-4{J^2 \over Q^3}\right]
\right)+\ldots ~,\cr M^I & = {q^I\over Q^{1/2}}\left( 1  +
{c_2\cdot q\over 144 Q^{3/2}}\left[1-4{J^2 \over Q^3}\right]
\right) +{ 1 \over 8Q^{1/2}}\left( 1 - {4\over 3}{J^2\over
Q^3}\right)C^{IJ}c_{2J} +\ldots ~. }}

\subsec{The 4D-5D connection}
 One of the advantages in introducing  the Gibbons-Hawking coordinates \ba\
 is that they facilitate the comparison between 5D and 4D points of view.

To see how this works, start with the rotating black hole solution
presented above and then reorganize the metric into a form suitable
for KK reduction along $x^5$,
\eqn\bm{ ds^2 = - e^{-4\phi} \left( dx^5 + \cos\theta d\phi\  +
A^0_t dt \right)^2 + e^{2\phi}\left( gdt^2 - g^{-1} (d\rho^2 +
\rho^2d\Omega^2_2)\right)~. }
Our {\it ansatz} gives
\eqn\bn{\eqalign{ e^{-4\phi} & = e^{-2U}\rho\left( 1 -
{1\over\rho}e^{6U}\omega^2_5\right)\quad \Rightarrow  \quad
 \ell^2 \left( 1- {\hat J}^2\right)~,\cr
g^2 & = {e^{6U}\rho\over 1 - {1\over\rho} e^{6U} \omega_5^2}\quad
\Rightarrow\quad   {\rho^4\over\ell^6(1-{\hat J}^2)}~, \cr A^0_t & =
-{e^{6U}\omega_5\over \rho \left( 1 - {1\over\rho} e^{6U}
\omega^2_5\right)} \quad \Rightarrow\quad -{{\hat J}\over 1- {\hat
J}^2}{\rho\over\ell^3} ~. }}
The arrows implement   the near horizon limit where the metric
function takes the form \bg. Since \eqn\bna{ e^{-2\phi}g =
{\rho^2\over\ell^2}~, } we see that the 4D string metric has
$AdS_2\times S^2$ near horizon geometry with the $AdS_2$ and the
$S^2$ both having radii $\ell$. The 4D Einstein metric
\eqn\bo{
ds^2_{4E} =  gdt^2 - g^{-1} (d\rho^2 + \rho^2d\Omega^2_2)~,
}
describes an extremal black hole. The 4D matter fields are the
dilaton $\phi$, the KK gauge field $A^0$, and additional gauge
fields $A_4^I$ and scalars $a^I$  coming from the reduction of the
5D gauge field {\it via} the decomposition
\eqn\bp{\eqalign{ A^I  = e^{2U} M^I (dt+\omega)
 &= e^{2U} M^I  \left( 1 - \omega_5 A^0_t\right)dt  + e^{2U} M^I\omega_5
  \left(dx^5 + \cos\theta d\phi +A^0_t dt\right)\cr & = A_4^I+a^I
   \left(dx^5 + \cos\theta d\phi +A^0_t dt\right) ~.}}
The 4D point of view will play a central role in the following.

\newsec{Entropy of 5D spinning black holes (and black rings)}

In this section we compute the entropy of our black holes.  This
is most conveniently done via the entropy function approach
\SenWA, which essentially amounts to evaluating the Lagrangian
density on the near horizon geometry.   The one complication is
that the entropy function method assumes a gauge invariant
Lagrangian, whereas we have non-gauge invariant Chern-Simons terms
in the action.   The remedy for this is well known
\refs{\SahooVZ}: we should reduce the action to 4D, and then add a
total derivative term to the Lagrangian to cancel the non-gauge
invariant piece.  Applications of the entropy function to rotating
black holes can be found in
\refs{\SahooVZ,\AstefaneseiDD,\MoralesGM,\CardosoRG,\GoldsteinKM}.
In the last subsection we consider black rings; for previous work
on the entropy function for black rings see
\refs{\DabholkarZA,\GoldsteinKM,\CaiUW}.

\subsec{Near horizon geometry and the entropy function}

We first review the general procedure for determining the entropy
from the near horizon solution, mainly following \GoldsteinKM. The
general setup is valid for spinning black holes as well as black
rings.

The near horizon geometries of interest take the form of a circle
fibered over an AdS$_2 \times S^2$ base:
\eqn\ca{\eqalign{ds^2& = w^{-1}\Big[v_1 \Big(\rho^2 d\tau^2
-{d\rho^2 \over \rho^2}\Big) -v_2 (d\theta^2 +\sin^2 \theta
d\phi^2) \Big] -w^2 \Big(dx^5 + e^0 \rho d\tau + p^0 \cos \theta
d\phi\Big)^2~, \cr A^I & = e^I \rho d\tau+p^I\cos \theta
+a^I\Big(dx^5 + e^0 \rho d\tau + p^0 \cos\theta d\phi\Big)~, \cr
v&= -{1 \over 4 \Nc} M_I F^I~. }}
The parameters $w$, $v_{1,2}$, $a^I$ and all scalar fields are
assumed to be constant.     KK reduction along $x^5$ yields a 4D
theory on AdS$_2 \times S^2$.  The solution carries the magnetic
charges $p^I$, while $e^I$ denote electric potentials.\foot{An
important point, discussed at length below, is that $e^I$ are
conjugate to 4D electric charges, which differ from the 5D
charges.}

Omitting  the Chern-Simons terms for the moment, let the action be
\eqn\cb{ I = {1 \over 4\pi^2}\int\! d^5x \sqrt{g}{\cal L}~. }
Define
\eqn\cc{ f= {1 \over 4\pi^2}\int\! d\theta d\phi dx^5
\sqrt{g}{\cal L}~. }
Then the black hole entropy is
\eqn\cd{ S= 2\pi \Big(e^0 {\p f \over \p e^0} + e^I {\p f \over \p
e^I} -f \Big)~. }
Here $w$, $v_{1,2}$ etc. take their on-shell values.  One way to
find these values is to extremize $f$ while holding fixed the
magnetic charges and electric potentials.  The general
extremization problem would be quite complicated given the
complexity of our four-derivative action.   Fortunately, in
the cases of interest we already know the values of all fields
from the explicit solutions.

The Chern-Simons term is handled by first reducing the action
along $x^5$ and then adding a total derivative to ${\cal L}$ to
restore gauge invariance.

\subsec{Computation of the on-shell action}
Starting from our solution written in the form \bm\ we insert the near
horizon values given in \bn\ and then change coordinates $t=\tau\ell^3\sqrt{1-{\hat J}^2}$
so that the
solution takes the form \ca. We then read off the magnetic charges
$p^0=1~,p^I=0$ and the electric fields
\eqn\cea{ e^0 = -{{\hat J} \over \sqrt{1-{\hat J}^2}}~,\quad e^I= {\Mh^I
\over 2\sqrt{1-{\hat J}^2} }~.}
Expressing the remaining quantities in terms of $e^{0,I}$ and $w=\ell\sqrt{1-{\hat J}^2}$
we find
\eqn\ce{\eqalign{v_1 =v_2\equiv V &=[1+(e^0)^2]w^3~, \cr a^I &
=-{e^0 e^I \over [1+(e^0)^2]}~,\cr M^I& = {1 \over w} { e^I \over
[1+(e^0)^2]}~, \cr v& = {3 \over 4}w d\tau \wedge d\rho -{3 \over 4}
w e^0 \sin\theta d\theta \wedge d\phi~, \cr D& = -{3 \over
w^2}{[1-(e^0)^2]\over [1+(e^0)^2]^2 }~. }}

The Gibbons-Hawking coordinates \ba\ have the periodicity $ x^5
\cong x^5 +4\pi$, so that \cc\ becomes
\eqn\cf{ f = {4V^2 \over w}  {\cal L}~.}
To proceed we need to evaluate the various terms in ${\cal L}$
using \ce.

\vskip.2cm

\noindent {\bf Two-derivative gauge invariant contribution:}

\vskip.2cm

As we have emphasized, the Chern-Simons terms require  special
considerations because they are not gauge invariant. The remaining
terms in the two-derivative action are
\eqn\cg{\eqalign{{\cal L}^{(2)}_{GI} =& -{1 \over
2} D -{3 \over 4}R +v^2 + \Nc ({1 \over 2} D-{1 \over 4}R +3 v^2)
+2 \Nc_I v^{ab} F^I_{ab} +{1 \over 4} \Nc_{IJ} F^I_{ab} F^{Jab}~. }}
Inserting the {\it ansatz} \ca\ with the relations \ce\ we find
\eqn\ch{f^{(2)}_{GI} = 4 {[1-(e^0)^2] \over [1+(e^0)^2]^3 }\cdot{1 \over 6} c_{IJK}e^I e^J e^K~.}

\vskip.2cm

\noindent{\bf Two-derivative Chern-Simons term:}

\vskip.2cm

We next turn to the special treatment needed for the gauge Chern-Simons term
\eqn\cj{ I_{CS}= {1 \over 24\pi^2 }\int c_{IJK} A^I\wedge F^J
\wedge F^K~.}
The reduction to 4D amounts to the decomposition
\eqn\ck{ A^I = A_4^I + a^I (dx^5 +A_4^0)~. }
If we simply insert this into \cj\ the resulting action has the form
\eqn\cl{ c_{IJK} A^I \wedge F^J \wedge F^K  = 2 c_{IJK} A^I_4
\wedge (F^J_4 + a^J F^0_4)\wedge da^K \wedge dx^5~~+~~{\rm
gauge~invariant~},}
where the first term is not gauge invariant because $A_4^I$
appears by itself  rather than as part of the field strength. The
 remedy for this is to redefine our original action by the addition of a total
derivative
\eqn\cm{ I_{CS}\Rightarrow I'_{CS} = {1 \over 24\pi^2 }\int
c_{IJK} \Big( A^I\wedge F^J \wedge F^K + d\Big[ A^I_4 \wedge (2
F^J_4 + F^0_4 a^J)a^K\wedge  dx^5 \Big]\Big) ~. }
This new action is not meant to replace our original 5D action in
general, but it  is the correct action to use in the 4D entropy
function because it is gauge invariant. It is now straightforward
to compute
\eqn\cn{ f^{(2)}_{CS} = {4(e^0)^2 (3 + (e^0)^2)
\over (1 + (e^0)^2)^3} \cdot{1 \over 6} c_{IJK}e^I e^J e^K~.}

\vskip.2cm

\noindent{\bf Four-derivative  gauge invariant contribution:}

\vskip.2cm

We next turn to the higher derivative terms in the action. Again, the Chern-Simons term
requires special consideration. Putting that term aside we have the action
\eqn\co{\eqalign{{\cal L}^{(4)}_{GI}& = {c_{2I} \over 24}
\Big( {1 \over 8}M^I C^{abcd}C_{abcd} +{1 \over 12}M^I D^2 +{1
\over 6}F^{Iab}v_{ab}D \cr &+{1 \over 3}M^I C_{abcd}
v^{ab}v^{cd}+{1\over 2}F^{Iab} C_{abcd} v^{cd} +{8\over 3}M^I
v_{ab} \hat{\cal D}^b \hat{\cal D}_c v^{ac} \cr &+{4\over 3} M^I
{{\cal D}}^a v^{bc} {{\cal D}}_a v_{bc} + {4\over 3} M^I {{\cal
D}}^a v^{bc} {{\cal D}}_b v_{ca} -{2\over 3} M^I
\epsilon_{abcde}v^{ab}v^{cd}{{\cal D}}_f v^{ef}\cr &+ {2\over 3}
F^{Iab}\epsilon_{abcde}v^{cf} {{\cal D}}_f v^{de}
 +F^{Iab}\epsilon_{abcde}v^c_{~f}{{\cal D}}^d v^{ef}\cr
& -{4 \over 3}F^{Iab}v_{ac}v^{cd}v_{db}-{1 \over 3} F^{Iab}
v_{ab}v^2 +4 M^I v_{ab} v^{bc}v_{cd}v^{da}-M^I (v^2)^2\Big)~.}}
with
\eqn\cp{ v_{ab}\hat{\Dc}^b \hat{\Dc}_c v^{ac} = v_{ab} \Dc^b \Dc_c
v^{ac}-{2 \over 3} v^{ac} v_{cb} R_a^{~b}-{1 \over 12} v^2 R~. }
Inserting the {\it ansatz} \ca\ with the relations \ce\ we find
\eqn\cq{f_{GI}^{(4)} = -{1 \over 8} {[1+(e^0)^2+(e^0)^4]\over
[1+(e^0)^2]^3}c_{2I}e^I~,}
after algebra using MAPLE. It is worth noting that every term in the
action contributes to this result.

\vskip.2cm

\noindent{\bf Four-derivative  Chern-Simons term:}

\vskip.2cm

Finally we must consider the mixed gauge-gravitational
Chern-Simons term:
\eqn\crr{ I_{CS}= {1 \over 4\pi^2} {c_{2I} \over 24\cdot16} \int
\! d^5x \sqrt{g}~\epsilon_{abcde} A^{Ia} R^{bcfg} R^{de}_{~~fg}~.
}
Again we reduce to 4D variables by inserting the decomposition
\ck. Since there will be a term with $A_4^I$ appearing by itself
and not in a field strength, the result will not be gauge
invariant in 4D. After implementing the 4+1 split on the curvature
tensor and writing $\epsilon_{abcde}R^{bcfg}R^{de}_{~~fg}$ as a
total derivative the relevant term becomes
\eqn\cs{\eqalign{ I_{CS} = -{1 \over 4\pi^2} {c_{2I} \over 24\cdot
16}  w^2 &\int\! dx^5 \int\! d^4x \sqrt{-g_4} \epsilon_{ijkl}
A_4^{Ii} \nabla^l \Big( 2 F^0_{4mn} R^{jkmn}\cr &+{1 \over 2}w^2
F^{0jk}_4 F^0_{4mn} F^{0mn}_4 + w^2 F^0_{4mn}F^{0jm}_4 F^{0kn}_4
\Big)~+~{\rm gauge~invariant}~, }}
where indices are raised and lowered by the $AdS_2\times S^2$ metric
\eqn\ct{ds_4^2 ={V \over w} \Big[ \Big(\rho^2 d\tau^2 -{d\rho^2
\over \rho^2}\Big) - (d\theta^2 +\sin^2 \theta d\phi^2) \Big] ~.}
Also,  $\sqrt{-g_4}$ and $\epsilon_{ijkl}$ are defined with
respect to this metric.

We then cancel off the non-gauge invariant part by modifying \crr\ as
$I_{CS}~\Rightarrow I'_{CS} = I_{CS}+\Delta I_{CS}$, with
\eqn\cu{\eqalign{  \Delta I_{CS}= {1 \over 4\pi^2} {c_{2I} \over
24\cdot 16}   w^2\! \int\!dx^5 \int\! d^4x \sqrt{-g_4}
\epsilon_{ijkl} \nabla^l \Big[A^{Ii}_4 \Big( 2 F^0_{4mn}
R^{jkmn}+&{1 \over 2}w^2 F^{0jk}_4 F^0_{4mn} F^{0mn}_4 \cr & + w^2
F^0_{4mn}F^{0jm}_4 F^{0kn}_4 \Big)\Big] ~.}}
We now compute
\eqn\cv{\eqalign{f^{(4)}_{CS} &= -{1\over 16} {(e^0)^2[1-(e^0)^2]
\over[1+(e^0)^2]^3} c_{2I}e^I -{1 \over 48} {[2+5(e^0)^2]\over
[1+(e^0)^2]^2} c_{2I}e^I~, \cr & = -{1 \over 24} {[1+5
(e^0)^2+(e^0)^4]\over [1+(e^0)^2]^3} c_{2I}e^I~, }}
where in the top line we showed the separate contribution of
$I_{CS}$ and $\Delta I_{CS}$. Note that $\Delta I_{CS}$ is nonvanishing
even in the nonrotating case $e^0=0$.

\subsec{Computation of entropy}
Our final result for the  on-shell action $f$ is found by adding the contributions
determined in the previous subsection
\eqn\cw{ f =f^{(2)}_{GI}+f^{(2)}_{CS}+ f^{(4)}_{GI}+f^{(4)}_{CS}
={4\over [1+(e^0)^2]}\left({1\over 6}c_{IJK} e^I e^J e^K  -{1 \over 24} c_{2I}e^I\right) ~. }
The entropy \cd\ is
\eqn\cx{ S=2\pi\Big( e^0 {\p f\over \p e^0} +e^I  {\p  f\over \p
e^I} -f\Big)  = {16\pi \over [1+(e^0)^2]^2}\left( {1\over
6}c_{IJK} e^I e^J e^K+{1\over 24} (e^0)^2c_{2I}e^I\right)~. }
We can rewrite this in terms of rescaled moduli using \cea:
\eqn\cy{ S = 2\pi \sqrt{1-\hat{J}^2} \Big({1 \over 6} c_{IJK}
\hat{M}^I  \hat{M}^J \hat{M}^K +{1 \over 6} \hat{J}^2 c_{2I}
\hat{M}^I \Big)~. }
This is our final result for the entropy of the spinning black
hole,  expressed in terms of the near-horizon moduli.

We can also express the entropy in terms of the conserved charges. We first
use \bl\ to find an expression in terms of geometrical variables
\eqn\cya{S = 2\pi \sqrt{(2\ell)^6-J^2} \Big( 1   +{c_{2I}
M^I \over 48\ell^2}\Big)~,}
and then expand to first order in $c_{2I}$ using \ble\ to find
\eqn\cyb{ S=
2\pi \sqrt{Q^3-J^2} \Big( 1   +{c_{2}\cdot q\over 16}{Q^{3/2}\over
(Q^3-J^2)}  + \cdots\Big)~. }
This is our expression for the black
hole entropy as a function of charges.

The microscopic understanding of these black holes is quite
limited.  However, our formulae do agree with the microscopic
corrections to the entropy where such results are available
\refs{\VafaGR,\HuangSB}.   Note that these special cases do not
involve rotation, and amount to reproducing the ${c_{2I} \over 8}$
term in \bk.

\subsec{Black ring entropy}

The entropy computation we have presented for the spinning black
hole is readily modified  to the black ring.  So although black
rings are not the focus of the present work we make a detour to
present the relevant entropy formula.   Since we just use the
entropy function computed from the near horizon geometry we will
only be able to give a formula for the entropy in terms of the
electric potentials.  To express the entropy in terms of charges
requires more details of the full black ring solution than are
presently available.

For the black ring the near horizon solution is
\eqn\ea{\eqalign{ds^2& = w^{-1}V\Big[ \Big(\rho^2 d\tau^2
-{d\rho^2 \over \rho^2}\Big) - d\Omega^2 \Big] -w^2 \Big(dx^5 +
e^0 \rho d\tau \Big)^2~, \cr A^I & =-{1 \over 2} p^I \cos \theta
d\phi-{e^I \over e^0} dx^5 ~.}}
Further details of the solution follow from the fact that the near
horizon geometry is a magnetic attractor, as studied in \CastroSD.
The near horizon geometry is a product of a BTZ black hole and an
$S^2$,  and there is enhanced supersymmetry.  These
conditions\foot{which can also be verified  by extremizing the
full entropy function.} imply
\eqn\ec{\eqalign{M^I & = {p^I \over 2w e^0}~, \cr
V&=  w^3(e^0)^2~,\cr
D& = {3 \over   w^2 (e^0)^2}~,
\cr v&=-{3 \over 4}w e^0
\sin\theta d\theta \wedge d\phi~,   }}
as can be read off from \CastroSD.

The computation of the $f$ function now proceeds just as for the
rotating black hole.  The result is
\eqn\ed{f =f^{(2)}_{GI}+ f^{(2)}_{CS}+f^{(4)}_{GI}+f^{(4)}_{CS} =
-{1 \over  2 e^0}({1\over 6} c_{IJK} p^I p^J p^K +{1 \over 6} c_{2I}p^I) +2 {c_{IJK}e^Ie^J
p^K \over e^0}~,}
and the entropy is
\eqn\ee{ S =2\pi( e^0{\p f \over \p e^0}+e^I{\p f \over \p e^I}-f)
={2\pi\over e^0} \left({1\over 6} c_{IJK} p^I p^J p^K +{1 \over 6} c_{2I}p^I\right)~.  }

The entropy can also be expressed as
\eqn\ef{ S = (2-\Nc){A \over \pi}=(2-\Nc){A \over 4 G_5}~,}
where $A$ is the area of the event horizon. In the two-derivative
limit we have $\Nc=1$ and we recover the Bekenstein-Hawking
entropy.

As mentioned above, the final step is to trade $e^0$ for the
charges of the black ring, but for this one needs knowledge of
more than just the near horizon geometry.

\newsec{Comparison with topological strings, the 4D-5D connection, and all that}

In this section we discuss various interpretational  aspects and
the relation to previous work.

\subsec{Comparison with 4D black hole entropy from the topological
string}

The OSV conjecture relates the  free energy of the  topological
string  to the Legendre transform of the 4D black hole entropy
\OSV.  It has further been proposed that the OSV conjecture lifts
to five dimensions  \GuicaIG. It is instructive to compare this 5D
version of the OSV conjecture with our explicit computations.  Our
analysis has been at the level of the 1-loop correction to the
free energy, and at this level the OSV conjecture for the entropy
{\it by design} reproduces the known 1-loop correction for the 4D
black hole.  So from a logical standpoint, our comparison below
really refers to the relation between 4D and 5D black hole
entropy.  We nevertheless find it useful to cast the discussion in
the language of the OSV conjecture, although this is not strictly
necessary.

The one-loop free energy from the topological string is
\eqn\da{
{\cal F} = {i\over\pi\mu}\left( {1\over 6} c_{IJK} \phi^I \phi^J \phi^K - {\pi^2\over 6}c_{2I}\phi^I\right)
+{\rm c.c.} = -{1\over\pi^2}{{1\over 6}c_{IJK}\phi^I \phi^J \phi^K - {\pi^2\over 6}c_{2I}\phi^I
\over \left({{\rm Re} \mu\over 2\pi}\right)^2 +1}~,
}
where $\mu = {\rm Re} \mu - 2\pi i$. The relation to our notation is
\eqn\db{\eqalign{
{\rm Re} \mu  &  =  2\pi e^0 =  - {2\pi{\hat J}\over\sqrt{1-{\hat J}^2}}~,\cr
 \phi^I  & = 2\pi e^I = {\pi{\hat M}^I\over\sqrt{1-{\hat J}^2}}~, \cr
{\cal F} & = - 2\pi f~.
}}
With these identifications we see that the free energy from the
topological string \da\ agrees precisely with our $f$ function
\cw. The 5D OSV conjecture gives the entropy
 \eqn\dc{
 S = {\cal F} - \phi^I {\partial{\cal F}\over\partial\phi^I}-
 {\rm Re} \mu {\partial{\cal F}\over\partial{\rm Re} \mu}
 = {2\over \pi^2  \left( \left({{\rm Re} \mu\over 2\pi}\right)^2 +1\right)^2}
 \left(
 {1\over 6}c_{IJK}\phi^I \phi^J \phi^K + ({\rm Re} \mu)^2{c_{2I}\over 24}
  \phi^I\right)~.
 }
This agrees precisely with our result \cy\ for the entropy. Of
course this second agreement is not independent from the first, since we Legendre transform the same expression on the
two sides.

So far we expressed the free energy and the entropy as  functions
of the potentials. However, we are usually more interested in
these quantities written in terms of the conserved charges
$(J,q_I)$.  According to our explicit construction of the solution
the charges are related to rescaled potentials through \bk.
Rewriting in terms of the electric fields \cea\ and then using the
dictionary \db\ to the topological string we have
\eqn\dd{\eqalign{
q_I & = {{1\over 2}c_{IJK} \phi^J \phi^K - {\pi^2\over 6}c_{2I}
\over \pi^2 \left( 1 +  \left( {{\rm Re} \mu\over 2\pi}\right)^2\right)} + {1\over 24}c_{2I}~,\cr
J &= - {{1\over 3!}c_{IJK} \phi^I \phi^J \phi^K - {\pi^2\over 6}c_{2I} \phi^I \over \pi^3
\left( 1 +  \left( {{\rm Re} \mu\over 2\pi}\right)^2\right)^2}{{\rm Re} \mu\over 2\pi}
- {1\over 12\pi}c_{2I} \phi^I {1\over 1 + \left( {{\rm Re} \mu\over 2\pi}\right)^2}
{{\rm Re} \mu\over 2\pi}~.
}}
The 5D OSV conjecture \GuicaIG\ instead defines the charges as
\eqn\de{\eqalign{
{\bar q}_I & = - {\partial{\cal F}\over\partial\phi^I}
 = {{1\over 2}c_{IJK} \phi^J \phi^K - {\pi^2\over 6}c_{2I}
\over \pi^2 \left( 1 +  \left( {{\rm Re} \mu \over 2\pi}\right)^2\right)}~,
\cr
 {\bar J} & = - {\partial{\cal F}\over\partial{\rm Re} \mu}
 = -  {{1\over 3!}c_{IJK} \phi^I \phi^J \phi^K - {\pi^2\over 6}c_{2I} \phi^I \over \pi^3
\left( 1 +  \left( {{\rm Re} \mu \over 2\pi}\right)^2\right)^2}{{\rm Re} \mu \over 2\pi}~,
 }}
and these do {\it not} agree with our expressions \dd.

A consequence of this discrepancy is that our expression for the
entropy disagrees with that conjectured in  \GuicaIG\ when both
are written in terms of conserved charges. In the notation used
here the topological string gives\foot{Notation: $q^I_{\rm
here}=y^I_{\rm there}$. Also note that \GuicaIG\ introduce moduli
$Y^I$ which satisfy the tree-level special geometry condition even
when higher derivative corrections are taken into account so these
moduli are shifted relative to $M^I$ used here.} \eqn\df{ S= 2\pi
\sqrt{{\bar Q}^3-{\bar J}^2} \Big(1+{c_{2}\cdot {\bar q}\over
12{\bar Q}^{3/2}}  + \cdots\Big)~. } This does not take the same
form as our expression \cyb.

The discrepancy arises because the 4D-5D charge  map used in
\GuicaIG\ misidentifies the 5D charges. The charges we have been
using, $(J,q_I)$, are the 5D conserved charges as measured by
surface integrals at infinity. In contrast, the charges from the
topological string, $({\bar J},{\bar q}_I)$, are defined via the
4D effective theory. The black hole with the prescribed near
horizon geometry, and which asymptotes to  4D asymptotically flat
spacetime (times a circle), has a Taub-NUT base space. As we show
explicitly in the next two subsections, the Taub-NUT itself has a
delocalized contribution to the 4D charges. This contribution is
absent for the 5D black hole.

\subsec{Spinning black hole on a Taub-NUT base space: the
solution}

 In order to carry out the 4D-5D reduction explicitly we
now construct the spinning black hole on a Taub-NUT base space. To
do so we need to  generalize some previous results to the case of
a curved base space. Most of the analysis goes through essentially
unchanged, so we can be brief.

 The analysis  of the  Killing spinor equations is
unchanged except that derivatives on the base space now become
covariant. As a result \aa-\ae\ remain valid on the Taub-NUT base
space. Supersymmetry also demands that $d\omega$ is a self-dual
two-form on the base space. Finally, supersymmetry requires a
Killing spinor which is covariantly constant on the base-space. This
in turns implies that the base-space is hyper-K\"ahler and so also
Ricci-flat, with anti-self-dual Riemann tensor. Using this
information it is straightforward to generalize Gauss' law \ah,
\eqn\fd{{\nabla}^2\left[e^{-2U}M_I-{c_{2I}\over24}\left(3({\nabla}U)^2
-{1\over4}e^{6U}(d\omega)^2\right)\right] ={c_{2I}\over24\cdot
8}{R}^{ijkl}{R}_{ijkl}~.}
Indices are contracted with the four-dimensional base space
metric, and the Riemann tensor and derivatives are that of the base space.  We see
that the only change is the new contribution on the right hand
side.  This in turn comes from the $A\wedge \Tr R^2$ term in the
action, which represents a curvature induced charge density.

We first consider the case of a charge $p^0=1$ Taub-NUT space, and
then generalize to the case of general charge.  We write Taub-NUT
in Gibbons-Hawking form
\eqn\fa{d{s}^2_4={1 \over H^0(\rho)}(dx^5+ \cos \theta
d\phi)^2+H^0(\rho)\left(d\rho^2+\rho^2(d\theta^2+\sin^2\theta
d\phi^2)\right)~,}
%
%with $i=(\rho,\theta,\phi)$ and
with $x^5 \cong x^5 +4\pi$ and  orientation
$\epsilon_{\rho\theta\phi x_5}=1$. The harmonic function $H^0$ is
\eqn\fb{H^0(\rho)=1+{1\over\rho}~.}
As in \bc, the anti-self-duality and closure conditions determine
$d\omega$ completely, {\it viz.}
\eqn\fc{d\omega=-{J\over 8 \rho^2}({e}^{\rho} e^5+e^\theta
e^\phi)~,}
where the $e^i$ are the obvious vielbeins of Taub-NUT.

For Taub-NUT the source on the right hand side of \fd\ can be
expressed as
\eqn\fe{{R}^{ijkl}{R}_{ijkl}=
{\nabla}^2\left({2\over\rho(\rho+1)^3}-{2 \over  \rho} \right)~.}
Using this we can easily solve \fd\ as
\eqn\ff{H_I = M_I^\infty + {q_I\over 4\rho} =
e^{-2U}M_I-{c_{2I}\over24}\left(3({\nabla}U)^2-e^{6U}{J^2\over64
\rho^4} +{1\over4}{1\over\rho(\rho+1)^3}-{1 \over 4 \rho}
\right)~,}
where we have substituted in \fc\ for $d\omega$. The radial
function $U(\rho)$ is determined again by the D equation of
motion which remains of the form \aj.

\subsec{Relation between 4D and 5D charges}

The above construction incorporates both 4D and 5D black holes.
Specifically, if we drop the $1$ in the harmonic function $H^0$
then the base space is simply $\IR^4$ and we recover the 5D black
hole.   Now, we have been using the symbol $q_I$, but we need to
check its relation to the physical electric charge of the 4D and
5D black holes.    From the gauge field dependent terms in the
action the conserved electric charge $Q_I$ is
\eqn\ffa{Q_I = -{1 \over 4\pi^2} \int_{\Sigma} \left( {1 \over 2}
{\cal N}_{IJ}  \star_5F^J +2 M_I  \star_5v \right)~,}
where $\Sigma$ denotes the $S^2\times S^1$ at infinity spanned by
$(\theta,\phi,x^5)$. Note that only the two-derivative terms in
the action contribute to \ffa\ since the four-derivative
contributions to the surface integral die off too quickly at
infinity.

Using the explicit solution we find
\eqn\ffb{ Q_I = -4\left[\rho^2 \p_\rho (M_I e^{-2U} )
\right]_{\rho = \infty}~.}
In the case of the 5D black hole we have $M_I e^{-2U} = H_I+
\ldots $, where $\ldots$ denote terms falling off faster than ${1
\over \rho}$, and hence we find
\eqn\ffc{ Q^{(5D)}_I = q_I~.}

For the 4D black hole we should instead use \ff, and we see that
the final term in parenthesis contributes an extra ${1 \over
\rho}$ piece.   Hence, for the 4D black hole we have  $M_I e^{-2U}
= H_I -{ c_{2I}  \over 4 \cdot 24 \rho} + \ldots $, which gives
\eqn\ffd{ Q^{(4D)}_I = q_I- {c_{2I} \over 24} = \bar{q}_I~.}

A similar story  holds for the relation between the 5D angular
momentum $J$ and the  4D charge $q_0$.   So we see that the 4D and
5D charges are different.  This has important implications for the
4D-5D connection: it is {\it not} true that $
S_{5D}(J,q_I)=S_{4D}(q_0=J,q_I)$. Rather, one should first convert
from barred to unbarred charges in the 4D entropy formula before
writing the result for the 5D entropy.    In general, if we write
$({\bar J},{\bar q}_I)= (J+\Delta J,q_I+\Delta q_I)$,  then we
should instead use $S_{5D}(J,q_I)= S_{4D}(J+\Delta J,q_I+\Delta
q_I)$.

The physical reason for this is simple: due to higher derivative
effects the Taub-NUT space itself carries a delocalized charge.
The 4D black hole sees the charge as measured at infinity, while
the 5D black hole effectively sees the charge as measured near the
tip of Taub-NUT (since the 5D black hole is obtained by dropping
the $1$ in $H^0$).   To see how these two notions of charge are
related, we  define a $\rho$ dependent ``charge'' via the
left-hand side of \fd, \eqn\fg{ Q_I(\Sigma_{\rho}) = -{1\over
4\pi^2} \int_{\Sigma_{\rho}} \sqrt{h} n^\mu {\nabla}_\mu
\left\{e^{-2U}M_I-{c_{2I}\over24}\left(3({\nabla}U)^2-{1\over4}e^{6U}
(d\omega)^2\right)\right\} } where $\Sigma_\rho$ is surface of
constant $\rho$ with unit normal, $n^\mu$, and $h$ is the induced
metric on $\Sigma_\rho$. Because of the curvature term in \fd,
this quantity is dependent on $\rho$. The difference between the
charges at the center (5D) and at infinity (4D) is given by
integrating the right-hand side of \fd,
\eqn\fh{\eqalign{
Q_I(\Sigma_\infty)-Q_I(\Sigma_0) &= -{1\over 4\pi^2}
\int_{\Sigma_{\infty}-\Sigma_0} \sqrt{h} n^\mu {\nabla}_\mu
\left\{e^{-2U}M_I-{c_{2I}\over24}\left(3({\nabla}U)^2-{1\over4}
e^{6U}(d\omega)^2\right)\right\}~,\cr &=  -{1\over 4\pi^2}
\int_{{\cal M}} \sqrt{{g}} {\nabla}^2
\left\{e^{-2U}M_I-{c_{2I}\over24}\left(3({\nabla}U)^2-{1\over4}
e^{6U}(d\omega)^2\right)\right\}~,\cr &=-{1\over
4\pi^2}{c_{2I}\over24\cdot 8} \int_{{\cal
M}}{R}^{ijkl}{R}_{ijkl}~. }} For a 4D Ricci-flat manifold, the
Euler number is given by \eqn\fj{ \chi({\cal M}) =
{1\over32\pi^2} \int_{\cal M} R_{abcd}R^{abcd}~, } which for Taub-NUT gives $\chi=1$. Thus \eqn\fk{
Q_I(\Sigma_\infty)-Q_I(\Sigma_0) = -{c_{2I}\over24}~, }
which accounts for the relation between $\bar{q}_I$ and $q_I$.

We  emphasize again that charges are completely unambiguous in 5D.
Also, in 5D the asymptotic charge $Q_I(\Sigma_\infty)$ agrees with
the near horizon charge $Q_I(\Sigma_0)$ because the base is flat.
The nontrivial relation is between the 4D and 5D charges in the
presence of higher derivatives.

\subsec{Generalization to charge $p^0$}

We can easily  generalize the above to Taub-NUT with arbitrary
charge $p^0$. This is defined by taking a $Z_{p^0}$ orbifold of
the charge $1$ solution. We identify $x^5 \cong x^5 + {4\pi \over
p^0}$.   To keep the asymptotic size of the Taub-NUT circle fixed
we take $H^0 = {1 \over (p^0)^2} +{1 \over\rho}$, which is a
choice of integration constant.    Finally, to put the solution
back in standard form we define $(\tilde{x}^5 = p^0 x^5,
\tilde{\rho} ={1 \over p^0}\rho)$. The general charge $p^0$
solution then has (dropping the tildes)
\eqn\faa{\eqalign{d{s}^2_4&={1 \over H^0(\rho)}(dx^5+ p^0\cos \theta
d\phi)^2+H^0(\rho)\left(d\rho^2+\rho^2(d\theta^2+\sin^2\theta
d\phi^2)\right)~, \cr H^0(\rho)& =1+{p^0\over\rho}~,\cr H_I &=
M_I^\infty+ {q_I \over 4 \rho}
% \cr d\omega & ={J\over 8
%\rho^2}({e}^{\rho}\wedge e^5+e^\theta\wedge e^\phi)
~.}}
Again, $q_I$ is the 5D electric charge.  The 4D electric charge is
now
\eqn\fah{ \bar{q}_I = q_I - {c_{2I} \over 24 p^0}~.}

\subsec{Example: $K3\times T^2$}

We conclude the paper by making our formulae completely explicit
in the special case of $K3\times T^2$. In this case $c_{1ij} =
c_{ij}$, $i,j=2,\ldots 23$ are the only nontrivial intersection
numbers and $c_{2i}=0$, $c_{2,1}=24$  are the 2nd Chern-classes.

Our procedure instructs us to first find the hatted variables in
terms of conserved charges by inverting \bk. In the present case
we find
\eqn\ga{\eqalign{
{\hat M}^1 & = \sqrt{\half c^{ij} q_i q_j + {4J^2\over (q_1+1)^2}\over q_1 + 3}~,\cr
{\hat M}^i & = \sqrt{ q_1 + 3\over \half c^{ij} q_i q_j + {4J^2\over (q_1+1)^2}} c^{ij} q_j~,\cr
{\hat J} & = \sqrt{ q_1 + 3\over \half c^{ij} q_i q_j + {4J^2\over (q_1+1)^2}}{J\over q_1+1}~.
}}
All quantities of interest are given in terms of these  variables.
For example, the relation between 4D charges \de\ and 5D charges
\db\ is
\eqn\gb{\eqalign{ {\bar J} & = {q_1-1\over q_1+1} J~,\cr {\bar q}_1
& = q_1 -1~, \cr {\bar q}_i & = q_i~, }}
and the entropy as function of the conserved charges becomes
\eqn\gc{\eqalign{ S &= 2\pi \sqrt{ \half c^{ij} q_i q_j (q_1 + 3) -
{(q_1-1) (q_1+3)\over (q_1+1)^2}J^2}\cr & =
2\pi\sqrt{(\bar{q}_1+4)\left[{1\over2}c^{ij}\bar{q}_i\bar{q}_j-{1\over\bar{q}_1}\bar{J}^2\right]}
~. }}
In the special case of $K3\times T^2$ the charge  corresponding to
D2-branes wrapping $T^2$ is special, and it is apparently that
charge which undergoes corrections due to higher order
derivatives. The precise form of the corrections is reminiscent of
the shifts in level that are characteristic of
$\sigma$-models.\foot{For example, the $\sigma$-model of heterotic
string theory on the near horizon geometry $AdS_3\times S^3/Z_N$
with ${\bar q}_1$ units of B-flux has spacetime central charge $c=
6({\bar q_1} + 4)$ \KutasovZH\ in apparent agreement with \gc.}

Our formulae for general Calabi-Yau black holes are democratic between the
various charges.

\bigskip
\noindent {\bf Acknowledgments:} \medskip \noindent   The work of
PK and JD is supported in part by NSF grant PHY-0456200. The work
of FL and AC is supported  by DOE under grant DE-FG02-95ER40899.

\appendix{A}{Derivation of the spinning black hole}
In this appendix we show how to obtain rotating  black hole
solutions by imposing the Killing spinor equations and Maxwell
equations, including higher derivatives corrections. Our
conventions follow \CastroHC.

We consider M-theory compactified on a Calabi-Yau threefold with
intersection numbers, $c_{IJK}$, and second Chern class
coefficients, $c_{2I}$. The bosonic part of the action up to
four-derivative terms is given by \eqn\va{S = {1\over4\pi^2} \int
d^5 x \sqrt{g}\left({\cal L}_0 + {\cal L}_1\right)~, } where the
two-derivative Lagrangian is
\eqn\vb{\eqalign{{\cal L}_0=& -{1\over2}D-{3\over4}R+v^2+{\cal
N}\left({1\over2}D-{1\over4}R+3v^2\right)+2{\cal
N}_Iv^{ab}F^I_{ab}\cr &+{\cal
N}_{IJ}\left({1\over4}F^I_{ab}F^{Jab}+{1\over2}\partial_aM^I\partial^aM^J\right)+{1\over24}
c_{IJK}A^I_{a}F^{J}_{bc}F^{K}_{de}\epsilon^{abcde}~, }} and the
four-derivative Lagrangian is
\eqn\vc{\eqalign{{\cal L}_1& = {c_{2I} \over 24} \Big( {1\over 16}
\epsilon_{abcde} A^{Ia} C^{bcfg}C^{de}_{~~fg} + {1 \over 8}M^I
C^{abcd}C_{abcd} +{1 \over 12}M^I D^2 +{1 \over 6}F^{Iab}v_{ab}D \cr
&+{1 \over 3}M^I C_{abcd} v^{ab}v^{cd}+{1\over 2}F^{Iab} C_{abcd}
v^{cd} +{8\over 3}M^I v_{ab} \hat{\cal D}^b \hat{\cal D}_c v^{ac}
\cr &+{4\over 3} M^I {\hat{\cal D}}^a v^{bc} {\hat{\cal D}}_a v_{bc}
+ {4\over 3} M^I {\hat{\cal D}}^a v^{bc} {\hat{\cal D}}_b v_{ca}
-{2\over 3} M^I \epsilon_{abcde}v^{ab}v^{cd}{\hat{\cal D}}_f
v^{ef}\cr &+ {2\over 3} F^{Iab}\epsilon_{abcde}v^{cf} {\hat{\cal
D}}_f v^{de}
 +F^{Iab}\epsilon_{abcde}v^c_{~f}{\hat{\cal D}}^d v^{ef}\cr
& -{4 \over 3}F^{Iab}v_{ac}v^{cd}v_{db}-{1 \over 3}
F^{Iab} v_{ab}v^2 +4 M^I v_{ab} v^{bc}v_{cd}v^{da}-M^I
(v^2)^2\Big)~.}}
The double superconformal derivative of the auxiliary field has curvature contributions
\eqn\vca{v_{ab}\hat{{\cal D}}^b\hat{{\cal D}}_cv^{ac}=v_{ab}{\cal
D}^b{\cal
D}_cv^{ac}-{2\over3}v^{ac}v_{cb}R_{a}^{\phantom{a}b}-{1\over12}v_{ab}v^{ab}R~.}
The functions defining the scalar manifold are
\eqn\vd{{\cal
N}={1\over6}c_{IJK}M^IM^JM^K~,\quad {\cal N}_I=\partial_I{\cal
N}={1\over2}c_{IJK}M^JM^K~,\quad {\cal N}_{IJ}=c_{IJK}M^K~,}
where $I,J,K= 1, \ldots, n_V$.

We study supersymmetric configurations so we seek solutions in which both the fermion fields and their first variations under supersymmetry vanish. The supersymmetry variations of the fermions are
\eqn\ve{\eqalign{
\delta\psi_\mu&=\left({\cal D}_\mu+{1\over2}v^{ab}\gamma_{\mu ab}-{1\over3}\gamma_\mu\gamma\cdot
v\right)\epsilon=0~, \cr
\delta\Omega^{I}&=\left(-{1\over4}\gamma\cdot
F^I-{1\over2}\gamma^a\partial_aM^I-{1\over3}M^I\gamma\cdot v
\right)\epsilon=0~,\cr
\delta \chi &=\left(D-2\gamma^c\gamma^{ab}{\cal
D}_av_{bc}-2\gamma^a\epsilon_{abcde}v^{bc}v^{de}+
{4\over3}(\gamma\cdot v)^2\right)\epsilon=0~.
}}
We now examine the consequences of setting these variations to zero.

\subsec{The stationary background}
We begin by writing our metric ansatz
\eqn\wa{ds^2=e^{4U_1(x)}(dt+\omega)^2-e^{-2U_2(x)}dx^i dx^i~,}
where $\omega=\omega_i(x)dx^i$ and $i=1\ldots4$. The vielbeins are
\eqn\wab{e^{\hat{0}}=e^{2U_1}(dt+\omega)~,\quad
e^{\ih}=e^{-U_2}dx^i~,}
which give the following spin connections
\eqn\wac{\eqalign{\omega^{\ih}_{\phantom{i}\jh}&=
e^{-U_2}(\partial_jU_2e^{\ih}-\partial_iU_2e^{\jh})+{1\over2}e^{2U_1+2U_2}d\omega_{ij}e^{\hat{0}}~,\cr
\omega^{\hat{0}}_{\phantom{0}\ih}&=2e^{U_2}\partial_iU_1e^{\hat{0}}+{1\over2}e^{2U_1+2U_2}d\omega_{ij}e^{\jh}~,}}
with
\eqn\wad{d\omega=\partial_{[i}\omega_{j]}dx^i\wedge dx^j~.}
The Hodge dual on the base space is defined as
\eqn\wb{\star_4\alpha_{\ih\jh}={1\over2}\epsilon_{\ih\jh\kh\lh}\alpha^{\kh\lh}~,}
with $\epsilon_{\hat{1}\hat{2}\hat{3}\hat{4}}=1$. A 2-form on the
base space can be decomposed into self-dual and anti-self-dual
forms,
\eqn\wbb{\alpha=\alpha^++\alpha^-~,}
where $\star_4\alpha^{\pm}=\pm\alpha^{\pm}$. We will use this
decomposition for the spatial components of $d\omega$ and the
auxiliary 2-form $v_{ab}$
\eqn\wbc{\eqalign{v_{\ih\jh}&=v^+_{\ih\jh}+v^-_{\ih\jh}~,\cr
d\omega_{\ih\jh}&=d\omega^+_{\ih\jh}+d\omega^-_{\ih\jh}~.}}
For stationary solutions, the Killing spinor $\epsilon$ satisfies
the projection
\eqn\wbd{\gamma^{\hat{0}}\epsilon=-\epsilon~.}
Using $\gamma_{abcde}=\epsilon_{abcde}$ and \wbd, it is easy to show
that anti-self-dual tensors in the base space satisfy
\eqn\wbe{\alpha^{-\ih\jh}\gamma_{\ih\jh}\epsilon=0~.}
\subsec{Supersymmetry variations}
There are three supersymmetry constraints we need to solve.
Following the same procedure as in \CastroHC, we first impose a
vanishing gravitino variation,
\eqn\wca{\delta \psi_\mu=\left[{\cal
D}_\mu+{1\over2}v^{ab}\gamma_{\mu
ab}-{1\over3}\gamma_{\mu}\gamma\cdot v\right]\epsilon=0~.}
Evaluated in our background, the time component of equation \wca\
reads
\eqn\wcb{\left[\partial_t-e^{2U_1+U_2}\partial_iU_1\gamma_{\ih}-{2\over3}e^{2U_1}v^{\hat{0}\ih}\gamma_{\ih}
-{1\over4}e^{4U_1}d\omega_{\ih\jh}\gamma^{\ih\jh}-
{1\over6}e^{2U_1}v_{\ih\jh}\gamma^{\ih\jh}\right]\epsilon=0~,}
where we used the projection \wbd. The terms proportional to
$\gamma_{\ih}$ and $\gamma_{\ih\jh}$ give the conditions
\eqn\wcc{\eqalign{v_{\hat{0}\ih}&={3\over2}e^{U_2}\partial_iU~,\cr
v^+&=-{3\over4}e^{2U_1}d\omega^+~.}}
The spatial component of the gravitino variation \wca\ simplifies to
\eqn\wcd{\left[\partial_i+{1\over2}\partial_jU_2\gamma_{\ih\jh}
+v^{\hat{0}\kh}e^{\jh}_i\left(\gamma_{\jh\kh}-{2\over3}\gamma_{\jh}\gamma_{\kh}\right)
-e^{\kh}_{~i}\left(v^{-}_{\kh\jh}+{1\over4}e^{2U_1}d\omega^-_{\kh\jh}\right)\gamma^{\jh}\right]\epsilon=0~,}
where we used the results from \wcc. The last term in \wcd\ relates
the anti-self-dual pieces of $v$ and $d\omega$,
\eqn\wce{v^-=-{1\over4}e^{2U_1}d\omega^-~.}
The remaining components of \wcd\ impose equality of the two
metric functions $U_1=U_2\equiv U$ and determine the Killing spinor as
\eqn\wcf{\epsilon=e^{U(x)}\epsilon_0~,}
with $\epsilon_0$ a constant spinor.

The gaugino variation is given by
\eqn\wda{\delta\Omega^I=\left[-{1\over4}\gamma\cdot
F^I-{1\over2}\gamma^a\partial_aM^I-{1\over3}M^I\gamma\cdot
v\right]\epsilon=0~.}
This constraint will determine the electric and self-dual pieces of
$F^I_{ab}$. Using \wbd\ and \wbe\ to solve \wda\ we find
\eqn\wdb{\eqalign{F^{I\hat{0}\ih}&=e^{-U}\partial_i(e^{2U}M^I)~,\cr
F^{I+}&=-{4\over3}M^Iv^+~. }}
Defining the anti-self-dual form
\eqn\wdc{\Theta^I=-e^{2U}M^Id\omega^-+F^{I-}~,}
then the field strength can be written as
\eqn\wdd{F^I=d(M^Ie^{\hat{0}})+\Theta^I~.}
We emphasize that $\Theta^I$, or more precisely $F^{I-}$, is undetermined
by supersymmetry. These anti-self-dual components are important for
black ring geometries but for rotating black holes we can take
$\Theta^I=0$ and $d\omega^-=0$.

Finally, the variation of the auxiliary fermion is
\eqn\we{\delta\chi=\left[D-2\gamma^c\gamma^{ab}{\cal
D}_av_{bc}-2\gamma^a\epsilon_{abcde}v^{bc}v^{de}+{4\over3}(\gamma\cdot
v)^2\right]\epsilon=0~.}
For the background given in section A.1 and using equations \wcc\
and \wce, the terms proportional to one or two gamma matrices cancel identically. The terms independent of $\gamma_{\ih}$ give an
equation for $D$, which reads
\eqn\wea{D=3e^{2U}(\nabla^2U-6(\nabla
U)^2)+{1\over2}e^{4U}(3d\omega^+_{\ih\jh}d\omega^{+\ih\jh}
+d\omega^-_{\ih\jh}d\omega^{-\ih\jh})~.}
\subsec{Maxwell equation}
The part of the action containing the gauge fields is
\eqn\xaa{S^{(A)}={1\over 4\pi^2}\int d^5x\sqrt{g}\left({\cal L}_0^{(A)}+{\cal L}_1^{(A)}\right)~,}
where the two-derivative terms are
\eqn\xa{{\cal L}_0^{(A)}=2{\cal N}_Iv^{ab}F^I_{ab}+{1\over4}{\cal
N}_{IJ}F^I_{ab}F^{Jab}+{1\over24}c_{IJK}A^I_aF^J_{bc}F^K_{de}\epsilon^{abcde}~,}
and the four-derivative contributions are
\eqn\xb{\eqalign{{\cal
L}_1^{(A)}={c_{2I}\over24}\bigg(&{1\over16}\epsilon^{abcde}
A^I_aC_{bc}^{\phantom{bc}fg}C_{defg}
+{2\over3}\epsilon_{abcde}F^{Iab}v^{cf}{\cal
D}_fv^{de}+\epsilon_{abcde}F^{Iab}v^{c}_{~f}{\cal
D}^dv^{ef}\cr&+{1\over6}F^{Iab}v_{ab}D
+{1\over2}F^{Iab}C_{abcd}v^{cd}-{4\over3}F^{Iab}v_{ac}v^{cd}v_{db}
-{1\over3}F^{Iab}v_{ab}v^2\bigg)~.}}
Variation of \xaa\ with respect to $A^I_{\mu}$ gives,
\eqn\xc{\eqalign{\nabla_\mu\left(4{\cal N}_Iv^{\mu\nu}+{\cal
N}_{IJ}F^{J\mu\nu}+2{\delta {\cal L}_1\over \delta
F^I_{\mu\nu}}\right)&\cr={1\over8}c_{IJK}
F^J_{\alpha\beta}F^K_{\sigma\rho}\epsilon^{\nu\alpha\beta\sigma\rho}
&+{c_{2I}\over24\cdot16}\epsilon^{\nu\alpha\beta\sigma\rho}
C_{\alpha\beta\mu\gamma}C_{\sigma\rho}^{\phantom{\sigma\rho}\mu\gamma}~,
}}
with
\eqn\xca{\eqalign{2{\delta {\cal L}_1\over \delta
F^{Iab}}={c_{2I}\over24}\bigg(&{1\over3
}v_{ab}D-{8\over3}v_{ac}v^{cd}v_{db}-{2\over3}v_{ab}v^2
+C_{abcd}v^{cd}\cr&+{4\over3}\epsilon_{abcde}v^{cf}{\cal
D}_fv^{de}+2\epsilon_{abcde}v^c_{~f}{\cal D}^dv^{ef}\bigg)~,}}
and
\eqn\xcb{{\delta {\cal L}_1\over \delta
F^I_{\mu\nu}}=e_{a}^{~\mu}e_{b}^{~\nu}{\delta {\cal L}_1\over \delta
F^I_{ab}}~.}

The equations of motion are evidently rather involved, so we will
now restrict attention to rotating black hole solutions with
\eqn\xd{d\omega=d\omega^+~,\quad d\omega^-=0~,\quad \Theta^I=0~.}
Given the form of the solution imposed by supersymmetry
it can be shown that the spatial components of the Maxwell equation are
satisfied automatically. The time-component of \xc\ give a non-trivial relation
between the geometry of the rotating black hole and the conserved
charges. We start by writing this equation as
\eqn\xda{\eqalign{\nabla_i\big(e^{-3U}[4{\cal
N}_Iv^{\ih\hat{0}}&+{\cal
N}_{IJ}F^{J\ih\hat{0}}]\big)+\nabla_i\left(2e^{-3U}{\delta {\cal
L}_1\over \delta F^I_{\ih\hat{0}}}\right)
-2e^{-2U}d\omega_{\ih\jh}{\delta {\cal L}_1\over \delta
F^I_{\ih\jh}} \cr&=e^{-4U}{1\over8}c_{IJK}
F^J_{ab}F^K_{cd}\epsilon^{\hat{0}abcd}
+e^{-4U}{c_{2I}\over24\cdot16}\epsilon_{\hat{0}abcd}C^{abfg}
C^{cd}_{\phantom{cd}fg}~.}}
The two-derivative contribution to \xda\ is
\eqn\xdb{\nabla_i\big(e^{-3U}[4{\cal N}_Iv^{\ih\hat{0}}+{\cal
N}_{IJ}F^{J\ih\hat{0}}]\big)-e^{-4U}{1\over8}c_{IJK}
F^J_{ab}F^K_{cd}\epsilon^{\hat{0}abcd}=-\nabla^2(e^{-2U}M_I)~,}
where we used the results from section A.2 and \xd. The higher
derivatives terms in \xda\ on this background are
\eqn\za{2{\delta {\cal L}_1\over \delta
F^{I\hat{0}\ih}}={c_{2I}\over24}e^{3U}\left(3\nabla_i(\nabla
U)^2-{9\over32}\nabla_i\left[e^{6U}(d\omega)^2\right]-{3\over8}e^{6U}\nabla_iU(d\omega)^2\right)}
\eqn\zab{-e^{-2U}d\omega_{\ih\jh}{\delta {\cal L}_1\over \delta
F^I_{\ih\jh}}={c_{2I}\over24}{3\over16}e^{6U}
\left(\nabla_kU\nabla_k(d\omega)^2+{1\over4}e^{6U}((d\omega)^2)^2+3(d\omega)^2\nabla^2U\right)}
\eqn\zac{\eqalign{e^{-4U}\epsilon_{\hat{0}abcd}C^{abfg}C^{cd}_{\phantom{cd}fg}=
&-{1\over2}\nabla^2[e^{6U}(d\omega)^2]+{3\over4}e^{12U}((d\omega)^2)^2\cr&+3e^{6U}(\nabla^2U-12(\nabla
U)^2)(d\omega)^2-3e^{6U}\nabla_kU\nabla_k(d\omega)^2
 }}

where again we used the form of the solution imposed by
supersymmetry and also the self-duality condition of $d\omega$.
Inserting \xdb-\zac\ in \xda\ gives
\eqn\zb{\nabla^2\left[e^{-2U}M_I-{c_{2I}\over8}\left((\nabla
U)^2-{1\over12}e^{6U}(d\omega)^2\right)\right]=0~.}
This is the generalized Gauss' law given in \ah.

 \listrefs
\end

\eqn\za{2{\delta {\cal L}_1\over \delta
F^{I\hat{0}\ih}}={c_{2I}\over24}e^{3U}\left(3\nabla_i(\nabla
U)^2-{9\over32}\nabla_i\left[e^{2U}(d\omega)^2\right]
-{3\over8}e^{2U}\nabla_iU(d\omega)^2\right)~,}
\eqn\zab{-e^{4U}d\omega_{\ih\jh}{\delta {\cal L}_1\over \delta
F^I_{\ih\jh}}={c_{2I}\over24}{3\over16}e^{2U}
\left(\nabla_kU\nabla_k(d\omega)^2+{1\over4}e^{2U}((d\omega)^2)^2
+(3\nabla^2U-4(\nabla U)^2)(d\omega)^2\right)~,}
\eqn\zac{\eqalign{e^{-4U}\varepsilon_{\hat{0}abcd}C^{abfg}C^{cd}_{\phantom{cd}fg}=
-&{1\over2}\nabla^2[e^{2U}(d\omega)^2]
+{1\over8}e^{4U}((d\omega)^2)^2-36e^{2U}(\nabla
U)^2(d\omega)^2\cr&-3e^{2U}\nabla_kU\nabla_k(d\omega)^2+3e^{2U}(\nabla^2U-4(\nabla
U)^2)(d\omega)^2~, }}
